\documentclass[12pt]{iopart}

\usepackage{graphicx}
\usepackage[bookmarks=true,colorlinks=true,linkcolor=blue]{hyperref}

\newcommand{\Eqn}[1]{Eq.~(\ref{#1})}

\newcommand{\Sec}[1]{Section~\ref{#1}}
\newcommand{\Secs}[1]{Sections~\ref{#1}}
\newcommand{\Fig}[1]{Fig.~\ref{#1}}
\newcommand{\Figs}[1]{Figs.~\ref{#1}}
\newcommand{\avg}[1]{\left\langle #1\right\rangle }
\newcommand{\triad}{\ensuremath{\vec{\mathcal{{T}}}}}

\begin{document}

\title[Synchronization and extinction in SIRS]{Synchronization and extinction in a high-infectivity spatial
  SIRS with long-range links}

\author{Ezequiel Arceo-May and Cristian F.~Moukarzel}

\address{CINVESTAV del IPN,  Appl.~Phys.~Dept.\\
97310 M\'erida, Yucat\'an, M\'exico.}
\ead{cristian.moukarzel@cinvestav.mx}



\begin{abstract}
  A numerical study of synchronization and extinction is done for a
  SIRS model with fixed infective and refractory periods, in the
  regime of high infectivity, on one- and two-dimensional networks for
  which the connectivity probability decays as $r^{-\alpha}$ with
  distance. In both one and two dimensions, a long-lasting
  synchronized state is reached when $\alpha < d$ but not when $\alpha
  > d$. Three dynamical stages are identified for small $\alpha$,
  respectively: a short period of initial synchronization, followed by
  a long oscillatory stage of random duration, and finally a third
  phase of rapid increase in synchronization that invariably leads to
  dynamical extinction. For large $\alpha$, the second stage is not
  synchronized, but is instead a long-lasting endemic state of
  incoherent activity. Dynamical extinction is in this case still
  preceded by a short third stage of rapidly intensifying synchronized
  oscillations.  A simple model of noise-induced escape from a
  potential barrier is introduced, that explains the main
  characteristics of the observed three-stage dynamical structure
  before extinction. This model additionally provides specific
  predictions regarding the size-scaling of the different timescales
  for the observed dynamical stages, which are found to be consistent
  with our numerical results.
\end{abstract}



\maketitle
\section{Introduction}
\label{chapter:sirs-on-sw}
Excitable systems~\cite{pattern-in-excitable-media,chialvo-model-95} are
interacting arrays of simple units possessing an active and a dormant
state.  These units can be activated by contact with an already active
neighbor, thereafter decaying to the inactive state. This decay
proceeds in a stochastic or deterministic manner, within a
characteristic decay time. Once deactivated, the unit must spend a
given refractory time in the inactive state before it can be activated
again.

Networks of interacting excitable units have been proposed as models
of brain dynamics
\cite{pattern-in-excitable-media,excitable-neurons-ws,dressed-neurons,asynchronous-state-cortical-circuits-renart2010,decorrelated-neuronal-firing-cortical-circuit-ecker2010,asynchronous-cortical-activity-parga2013},
disease propagation
\cite{pattern-in-excitable-media,sirs-sw-ka}, and more
\cite{pattern-in-excitable-media}. Despite their apparent
simplicity, collective effects endow these systems with interesting
dynamical behaviors such as the appearance of waves, complex
spatiotemporal patterns \cite{pattern-in-excitable-media} and
spontaneous synchronization \cite{sync-strogatz}.

Recurrent diseases may develop into an \emph{endemic} stable state,
where active and inactive states randomly coexist, without any kind of
spatiotemporal order. The stable state is a fixed point of the
dynamics in the limit of large systems.  For finite populations,
however, stochastic fluctuations lead to disease extinction in finite
time~\cite{extinction-sir-removal-renewal,persistence-extinction-spatial-sirs}. This
is called extinction by chance, or spontaneous extinction. A different
mechanism for activity extinction occurs through extreme
synchronization~\cite{behaviors-sirs-sw,sirs-sw-ka}, in models that
display such dynamical behavior. This last case is the focus of the
present work.

Understanding the factors affecting the duration, or \emph{persistence
  time}, of e.g.~an epidemic outbreak, is clearly a matter of the
highest relevance in epidemiological modeling.  Also important are
synchronization effects in the case of recurrent diseases such as
Influenza A H1N1~\cite{mobility-travel-h1n1} and other influenza-like
diseases~\cite{viboud2006,multiscale-mobility,gleam-model,heterogeneity-mobility-influenza}.

Synchronization is perhaps of even greater relevance in neural
systems, where it is believed to play a role in cognitive
proceses~\cite{neuronal-communication-by-neuronal-coherence}.  Extreme
synchronization appears, on the other hand, to be functionwise
undesirable in neural systems. It is known that certain dysfunctional
behaviors such as epileptic
seizures~\cite{LBHKRSW09,non-linear-analysis-egg-meg,synch-applied-to-epileptic-signals}
and Parkinson's disease~\cite{nm-phase-locking-detection-noisy-data},
are correlated with a high level of synchronism in certain areas of
the brain.  In models of excitable systems, extreme synchronism (phase
ordering) implies the extinction of the dynamics, since significant
numbers of both active and inactive units must be simultaneously
present in order for activity to be propagated in time.

The extinction of one or more among $n$ interacting subpopulations can
be mapped onto the so-called ``Exit
Problem''~\cite{exit-problem-grasman99}, in which the vector
$\vec{\mathbf{x}}$ of densities evolves within a space $\Omega$, until it
gets trapped by an absorbing boundary $\partial \Omega$. In the
presence of stochasticity, it is helpful to think of the exit problem
as that of slow diffusion of particles in a deterministic flow.  The
scaling of average exit times $\avg{t_{exit}}$ will then depend on the
nature of the underlying flow. Two particular
cases~\cite{stochastic-des-schuss,exit-problem-grasman99} are relevant
for this work.  \emph{Diffusion along a flow}: occurs when the
deterministic dynamics pushes the system out from $\Omega$. The
expected exit time, in general dimension, behaves
as~\cite{exit-problem-grasman99}
\begin{equation}
  \avg{t_{exit}} \sim \ln(N).
  \label{eq:extinction-time-along-flow}
\end{equation}
\emph{Diffusion against a flow}: occurs when the deterministic
dynamics have a stable equilibrium in $\Omega$, so that escape occurs
against the deterministic forces. The expected exit time, in dimension
one, is~\cite{exit-problem-grasman99}
\begin{equation}
  \avg{t_{exit}} \sim N^{-1/2} \exp(Q N) \qquad Q > 0.
  \label{eq:extinction-time-against-flow}
\end{equation}	
van Herwaarden \& Grasman~\cite{extinction-sir-removal-renewal}
and Roozen~\cite{exit-problem-2d-roozen89} reported the same
relation as \Eqn{eq:extinction-time-against-flow} for the SIR model
with removal-renewal and the generalized Lotka-Volterra model,
respectively. 
	
In ecology studies, the above scaling relations are observed in
stochastic Lotka-Volterra models
\cite{cyclic-lotka-volterra,edge-neutral-evolution,general-lotka-volterra}
and in stochastic SIRS models
\cite{persistence-extinction-spatial-sirs,behaviors-sirs-sw}.

For each instance of the excitable dynamics, the \emph{extinction
  time} is a random variable, whose distribution equals minus the
time-derivative of the \emph{survival probability} or
\emph{persistence}, $F(t)$, estimated as the fraction of active
systems at time $t$ ~\cite{extinction-nassel-sir}. Understanding
extinction times is of practical importance e.g.~in the design of
epidemiological policies (See \cite{extinction-branching} and
references therein).

Ki Baek~\cite{behaviors-sirs-sw} mention that extinction may
happen \emph{spontaneously} (by chance), or \emph{synchronously}.  For
the SIRS model in small-world networks, Kuperman \&
Abramson~\cite{sirs-sw-ka} reported that high values of virulence
usually lead to \emph{synchronous extinctions}.

The ability of a system to spontaneously synchronize depends on the
dimension and other topological properties of the network of
interactions among units
\cite{comnet-bocalleti-2006,synch-comp-net}. In the case
of short-range interactions, the lower critical dimension for the
appearance of cooperative synchronization is $d_c=2$
\cite{entrainment-lower-critical-dimension,sakaguchi-kuramoto-model,synch-in-spatial-extended-oscillators}.
Chains of Kuramoto oscillators with long-range interactions decaying
as $1/r^{\alpha}$ do synchronize, however only for $\alpha \leq 3/2$
\cite{long-range-km-spin-wave}.

Recent work analyzing connections between
neurons~\cite{BDM190,SGLM191,MTRWG192,PBM193} suggests that, while most
of them are short-ranged~\cite{PYB196,BS197}, a significant number of
long-distance connections also exists~\cite{SYPB198}.

One common element to various types of analises~\cite{KPRL1,KH2,HCE3}
of brain networks is the fact that their topology is that of a
small-world network, either with power-law distributed link
lengths~\cite{KPRL1}, exponentially truncated~\cite{KH2}, or a
combination of both~\cite{HCE3}.  At the level of functional brain
networks, it has been argued that their topology has elements from
small-world and scale-free networks~\cite{PRLEguiluzChialvo} Human
mobility patterns~\cite{HMP06}, which are relevant for disease
propagation, also show a power-law distribution of travelled
distances.

A simple model of network connectivity that allows onw to consider
both short- and long-range links in a controlled manner is the
power-law decay network~\cite{CFM05}. In this model, which is the one
we chose to use in this work, two nodes $i$ and $j$ in a
$d$-dimensional array are connected at random with probability $p_{ij}
\sim 1/r_{ij}^{\alpha}$, where $r_{ij}$ is their Euclidean distance,
and the decay exponent $\alpha$ is a tunable parameter. The density of
links $p$ is an additional parameter, which must be chosen large
enough to ensure macroscopic connectedness across the system. For
large $\alpha$, only short-range links are present, and a
$d$-dimensional topology is obtained. When $\alpha \to 0$, on the
other hand, long links are generated and an $\infty$-dimensional
random graph is produced.  In this article, the effects of spatial
dimension $d$ and the distribution of link lengths parametrized by
$\alpha$ are analyzed for a SIRS
\cite{hethcote-sirs,sirs-sw-ka,behaviors-sirs-sw,persistence-extinction-spatial-sirs,sirs-delay}
model of excitable dynamics on these networks.  The focus of the
present work is on the analysis of synchronization and the
characterization of extinction times in these networks.

Our large-scale numerical simulations show that networks with $\alpha
<d$ display three dynamic stages, when starting from an incoherent
state of random activity. These are: 1) a short transient of rapidly
increasing synchronization, followed by 2) a relatively large period
of sustained moderate coherence, and finally, 3) a short burst of
rapidly increasing coherence, that invariably leads to extreme
synchronization and extinction.  When $\alpha < d$, we argue that the
extinction process can be seen as noise-mediated escape from a stable
periodic orbit. For large $\alpha$, on the other hand, there is no
synchronized second stage, and the extinction of the dynamics can be
understood as an extreme-synchronization mediated escape from a fixed
point.  Rationalizing the extinction problem as one of stochastic
escape from an attractive periodic orbit \emph{or} fixed point,
extinction times are expected to scale in a similar way as those for
escape against a flow mentioned above, i.e.~as $\log{(t_{1/2} N)} \sim
N$.  Our numerical measurements are found to be consistent with this
expectation.  In all cases the timescale for extinction is found to
grow exponentially fast with increasing system size $N$. Notice that,
because we work in the regime of large infectivity, our results apply
to the limit of weak noise. The large-noise case, in which spontaneous
extinction is predominant, may have different scaling
properties~\cite{persistence-extinction-spatial-sirs}.

The rest of this article is organized as follows:
\Sec{section:models-and-methods} describes the networks and dynamical
model we use, discussing details of implementation, numerical
simulation, and measurement procedures.

\Sec{section:simulation-results} presents our numerical
results, as follows: \Secs{sync} and \ref{latestage} describe the
observation and characterization of dynamical stages leading to
extinction.

\Sec{sec:extinction-as-escape} introduces our model for escape from a
periodic orbit. Its predictions are discussed and compared with
numerical results in \Secs{exp-dist-t2}, \ref{sec:scal-refr-time}, and
\ref{tau2scaling}. In \Sec{sec:speed-growth} the $\alpha$-dependence
of the involved timescales is discussed, analyzing evidence for a
dynamical phase transition that is seen in one dimension but not in
two. Finally, \Sec{sec:concl-disc} presents a general discssion of our
results.
\section{Model and Methods}
\label{section:models-and-methods}
\subsection{SIRS model}
\label{subsection:sirs-model}
SIRS is a simple model of excitable systems that is able to reproduce
some important characteristics of neuronal networks and recurrent
diseases, such as oscillations and spatial
waves~\cite{hethcote-sirs,sirs-delay}.  SIRS dynamics is defined, at
each timestep, by the following three transition rules:
\begin{itemize}
\item[$S \to I$] : a susceptible individual having $k$ infectious
  neighbors becomes infected with probability $p = 1 -
  \left(1-p_{0} \right)^{k},$ where $p_{0} \in [0,1]$ is the
  \emph{link infectivity}.
\item[ $I \to R$ ]: an infected individual becomes refractory exactly
  $\tau_{i}$ timesteps after being infected.
\item[ $R \to S$] : a refractory node becomes susceptible after
  $\tau_{r}$ time steps.
\end{itemize}
Only the $S \to I$ transition is stochastic and depends on the node's
environment. The other two transitions are deterministic and happen at
pre-specified times after infection. Because of this, at a given
simulation time, the state of node $i$ is entirely
determined~\cite{sirs-sw-ka} by the time $t_{i}$ of its last
infection, as follows: node $i$ is infectious if $0 \leq t - t_{i} <
\tau_{i}$, refractory if $\tau_{i} \leq t - t_{i} < \tau_{A} =
\tau_{i} + \tau_{r}$, and susceptible if $t - t_{i} \geq
\tau_{A}$. Here $\tau_{i}$, $\tau_{r}$ and $\tau_{A}$ are,
respectively, the \emph{infectious}, \emph{refractory}, and
\emph{active} periods.  Since the time that a node spends in the
susceptible state is a stochastic variable, SIRS behaves as a
collection of \emph{non-identical oscillators} with different
frequencies.

The phase $\theta$ of an active node $i$ is
defined~\cite{sirs-sw-ka} as \hbox{$\theta_{i} = 2 \pi
  (t-t_{i})/\tau_{A} $}. From these, a synchronization order parameter
is defined by \cite{kuramoto-model,kuramoto-crawford}
\begin{equation}
z e^{\mathrm{i} \Theta} = \frac{1}{N_{a}} \sum_{j=1}^{N_{a}}
  e^{\mathrm{i} \theta_{j}},
  \label{eq:1}
\end{equation}
where the modulus $z$ of the phase vector is the \emph{coherence},
$\Theta$ is the \emph{mean phase}, and $N_{a}$ the number of active
(i.e.~$I+R$) sites.  The system is initialized by assigning a
uniformly distributed infection time $t_{i} \in [0,\tau_{A}]$ to every
node $i$, and setting $t = \tau_{A}$.  To actually implement the SIRS
dynamics, we use a procedure that significantly speeds up numerical
simulation.  We begin by defining a node to be \emph{infectible} if:
a) it is susceptible and it has at least one infected neighbor, or b)
it is a refractory node with at least one infected neighbor that will
still be infectious when the refractory node becomes susceptible.
Infectible nodes constitute the \emph{surface} where the
non-deterministic part of the dynamics, i.e.~the $S \to I$
transitions, takes place. Equivalently, for each infected node $i$,
all its neighbors $j$ with \hbox{$t_{i}-t_{j} > \tau_{r}$} are
infectible.

In our simulations, and given the starting set of infection times
$\{t_i\}$, we first set up a queue containing infectible sites. Each
site $i$ in the infectible queue is considered in turn for stochastic
infection. If $i$ does not get infected, then it is resent to the end
of to the queue, unless all of its infected neighbors have already
healed, in which case we $i$ is just ignored. If on the other hand $i$
does become infected, we set its infection time $t_i=t$, and send its
infectible neighbors to the end of the queue, using the simple rule
mentioned above to identify them.  One sweep through the entire queue
of infectible nodes constitutes a time-step, after which the time
variable $t$ is advanced by one unit. The simulation ends when the
queue of infectible sites becomes empty.  This procedure is
significantly faster than visiting all nodes, since only infectible
nodes must be considered.  As a consequence, we obtain a speed gain of
about $\times 10$.

\subsection{Networks in $d$ dimensions with power-law distribution of link lengths}
\label{subsection:power-law-link-length-distribution}
We next describe how $d$-dimensional networks of $N$ nodes, randomly
connected via $M$ long-range connections, are constructed in this
work. We intend to build networks for which the probability that an
arbitrary pair $(i,j)$ is connected by a link behaves as
$p_{i,j} \propto r_{ij}^{-\alpha}$, where $r_{ij}$ is the Euclidean
distance between nodes $i$ and $j$. Here $\alpha$ is the
\emph{connectivity decay exponent}, or link-exponent, for
short. Starting from a set of nodes in $d$-dimensional space, when
$\alpha$ is zero a random graph is obtained, in which each pair of
sites is connected with equal probability $p_{ij}=2M/(N(N-1))$. For
large $\alpha$, on the other hand, only short-range links are
generated, and a $d$-dimensional network with connectivity disorder is
obtained.
\\
Among several possible numerical procedures to obtain the desired
dependence of link-connectivity on Euclidean distance $r$, we chose,
for practical reasons, the one that follows. Initially, $N = L^{d}$
nodes/sites are assigned to the vertices of a hyper-cubic
$d$-dimensional lattice of linear dimension $L$ without any links. For
coding convenience, helicoidal boundary conditions are used. We next
generate $M$ links with random lengths $\ell$, distributed according
to
\begin{equation}
p(\ell) = A \ell^{-\alpha+d-1} \qquad \ell_{min} \leq \ell \leq \ell_{max},
\label{eq:power-law-length-distribution}
\end{equation}
and use each of them to connect a randomly chosen pair of sites,
provided the distance between them is approximately
$\ell$. $\ell_{min} =1$ and $\ell_{max} \sim L/2$ are, respectively,
the lower and upper bounds for link lengths on a finite system.  This
produces a random network with approximately Poisson degree
distribution and average degree $\avg{k} = 2M/N$, in which the
connectivity probability decays as $r^{-\alpha}$. Notice the extra
$d-1$ in \Eqn{eq:power-law-length-distribution}, which accounts for
the number of sites at a distance $\ell$ from a given one in $d$
dimensions.  In this work, networks embedded in $d=1$ and $d=2$
dimensions were generated, using the procedure sketched above.
\\ Because of the sparsity and randomness of our networks, a large
number of connected components can in general coexist. However, a
large enough $M=2N$ is always used, $M/N = \avg{k}/2 = 2$, so that the
largest connected component typically contains a large fraction of the
system. In other words, we work well above the percolation critical
density of links, located at approximately $\avg{k}_{c} =
1$~\cite{er-oeorg}.  \\ For each value of $d$, $\alpha$ and $L$, we
performed $10^3$ repetitions of the dynamics until extinction on a
newly generated network. Simulation of systems with several values of
$\alpha$ for moderate size as $N = 40 \times 40$ up to extinction
required weeks of cpu time in total.  \\ Because of the numerical
difficulties associated with long extinction times on large static
networks, we devised an annealing procedure that, we found, speeds up
the dynamics of the sytem. We take a fixed small \emph{annealing
  probability} $p_{a}$ and, at each time step, reallocate a total of
$p_{a} \times M$ links (both edge-ends). Since link-lengths are
unchanged, this procedure produces a slightly different network with
the same length distribution as the original one. We found that the
synchronization and extinction process happens faster on annealed
networks than on static networks.  Static properties measured over
annealed networks with small values of $p_{a}$, however, were found to
present the same quantitative behavior as on static networks without
annealing. The dynamical behavior is also similar on annealed
networks, only on shorter timescales.
\subsection{Measurement procedure}
\label{subsection:measurement-procedure}
For each repetition of the experiment, which involves a newly
generated network, we determine the \emph{extinction time} $t_{ext}$,
i.e. the number of timesteps at which the queue of infectible sites
becomes empty. We find that both the average and dispersion of this
random variable increase exponentially fast when $L$ increases. We are
also interested in estimating the distribution of extinction times
$P(t_{ext})$. The cumulative distribution of extinction times is
related to the survival probability or \emph{persistence},
$F(t)=\int_t^{\infty} P(t_{ext}) dt_{ext}$, which we estimate as the
fraction of still active systems at time
$t$~\cite{extinction-nassel-sir}.  The persistence
\emph{half-time} $t_{1/2}$ is defined by the condition
$F(t_{1/2}) = \frac{1}{2}$, i.e.~the time at which roughly one half of
all starting configurations have become extinct.
\\
For most observables of interest we take measurements, and average
their results over repetitions of the experiment, at geometrically
increasing time intervals $t_k \sim a^k$. Additionally, a second
measurement protocol was implemented in order to study the final
stages of the dynamic. To this end, a circular array of ``diameter''
$10^4$ is kept, that saves results from the last $10^4$ timesteps. We
average these ``final stage'' data after each simulation ends, in such
a way that measurements at extinction coincide for different
repetitions of the experiment. This allows us to obtain averages that
are representative of typical behavior on approach to extinction. We
denote the time measured backwards from extinction,
\hbox{$t_{te} = t_{ext}-t$}, as \emph{time-to-extinction}.
\\
Averages were taken in all cases over 1000 networks, with one random
starting configuration per network, for each set of parameters.  Link
infectivity $p_{0}$ was fixed to 0.75, a relatively large value, which
makes spontaneous (i.e.~not mediated by extreme synchronization)
extinctions unlikely \cite{sirs-sw-ka}.  The infectible period
was set to $\tau_{i} = 4$, and the refractory period to
$\tau_{r} = 8$.
\section{Simulation results}
\label{section:simulation-results}
\subsection{Synchronization and its $\alpha$-dependence}
\label{sync}
\begin{figure}[h!]
  \centerline{
    \includegraphics[width=0.38\linewidth,angle=270]{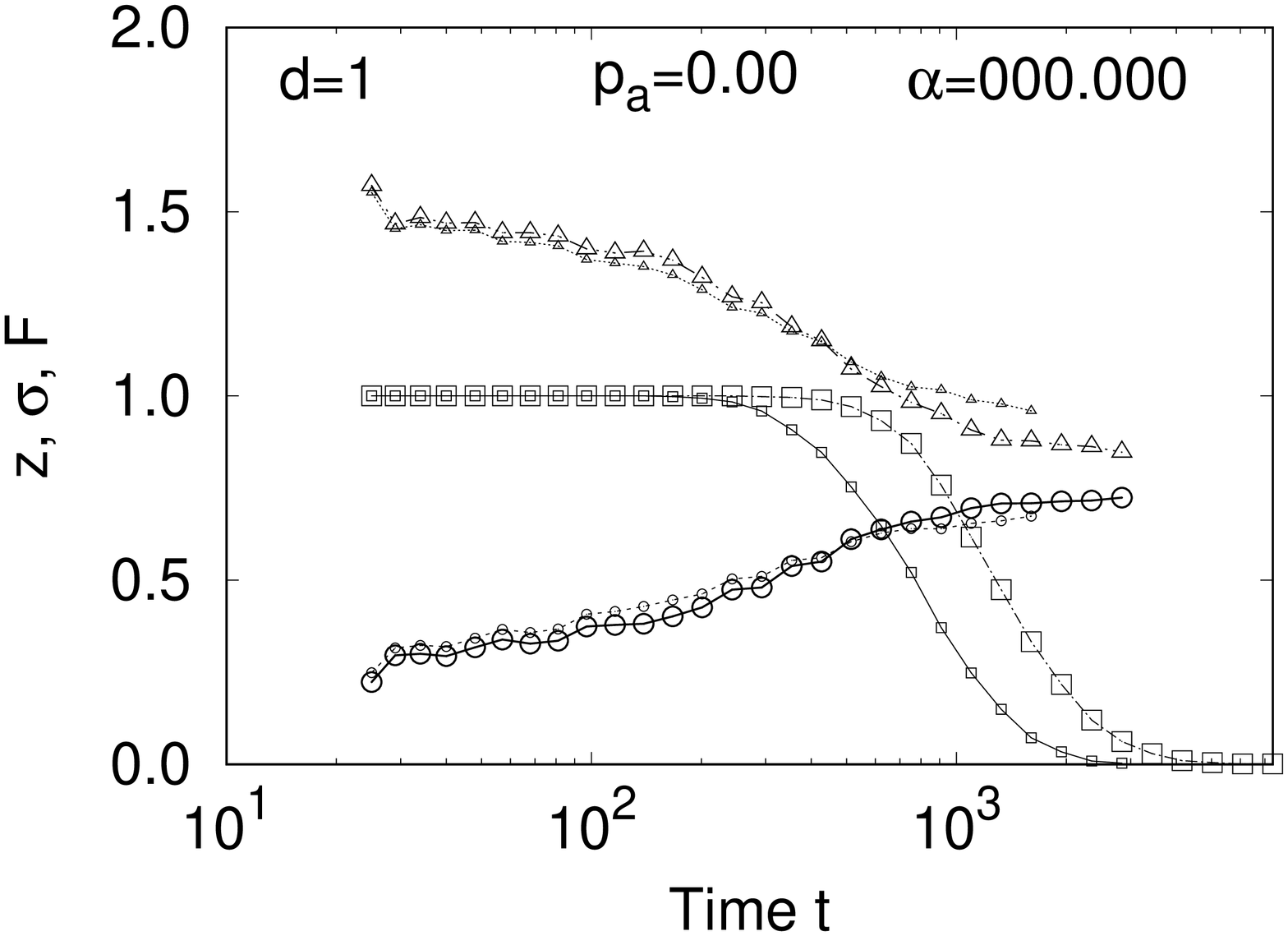}
    \includegraphics[width=0.38\linewidth,angle=270]{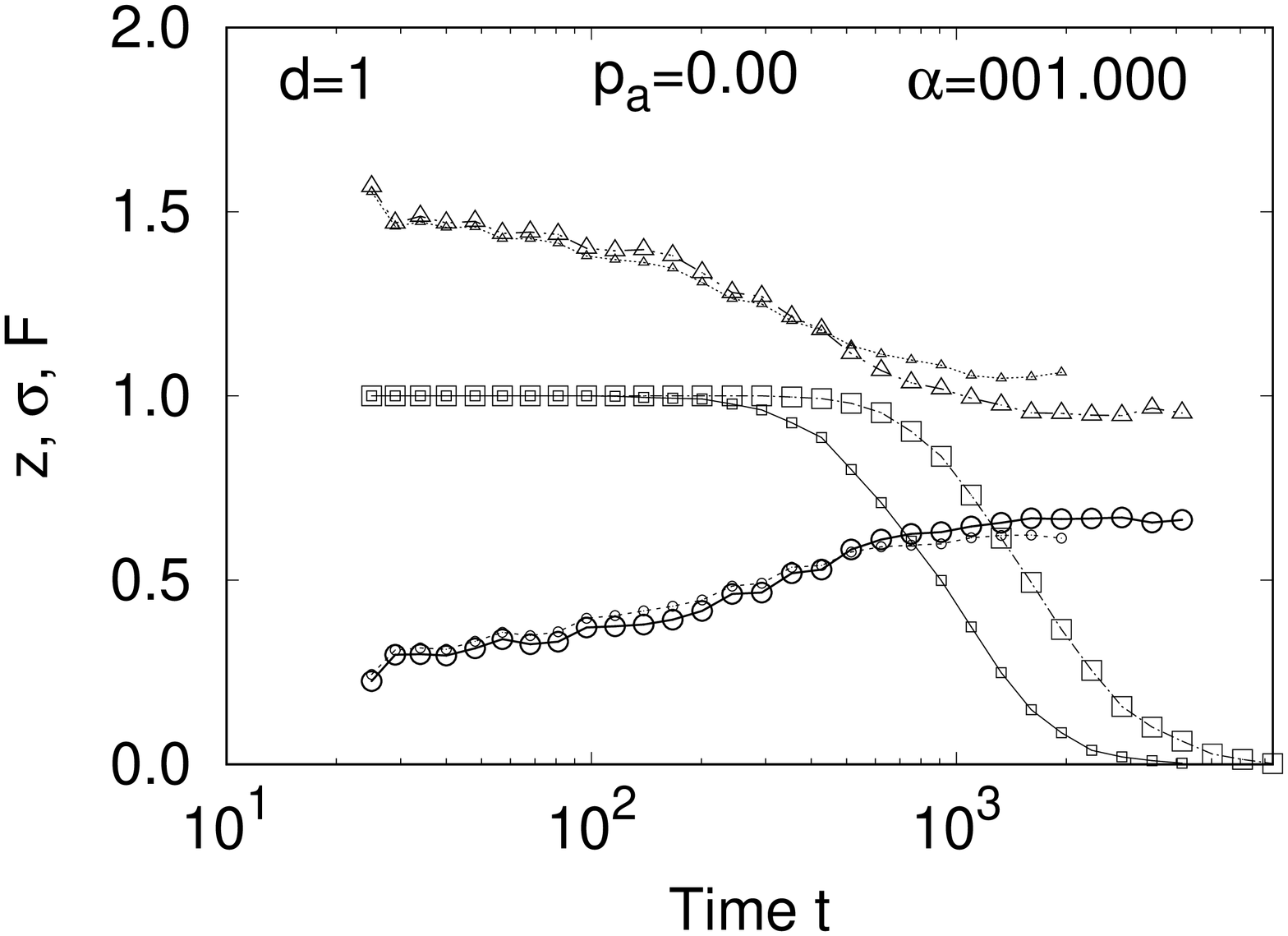}
  } \centerline{
    \includegraphics[width=0.38\linewidth,angle=270]{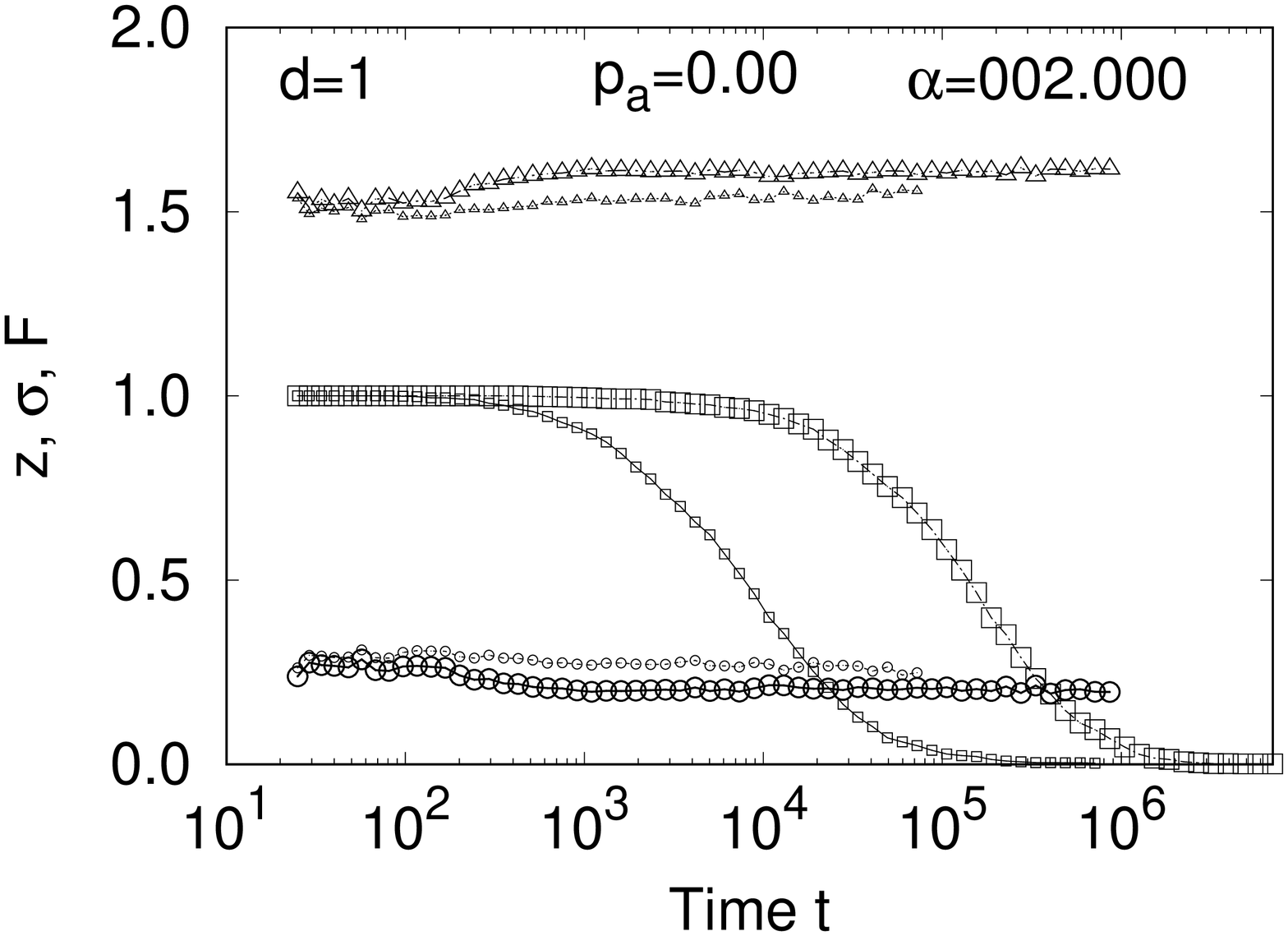}
    \includegraphics[width=0.38\linewidth,angle=270]{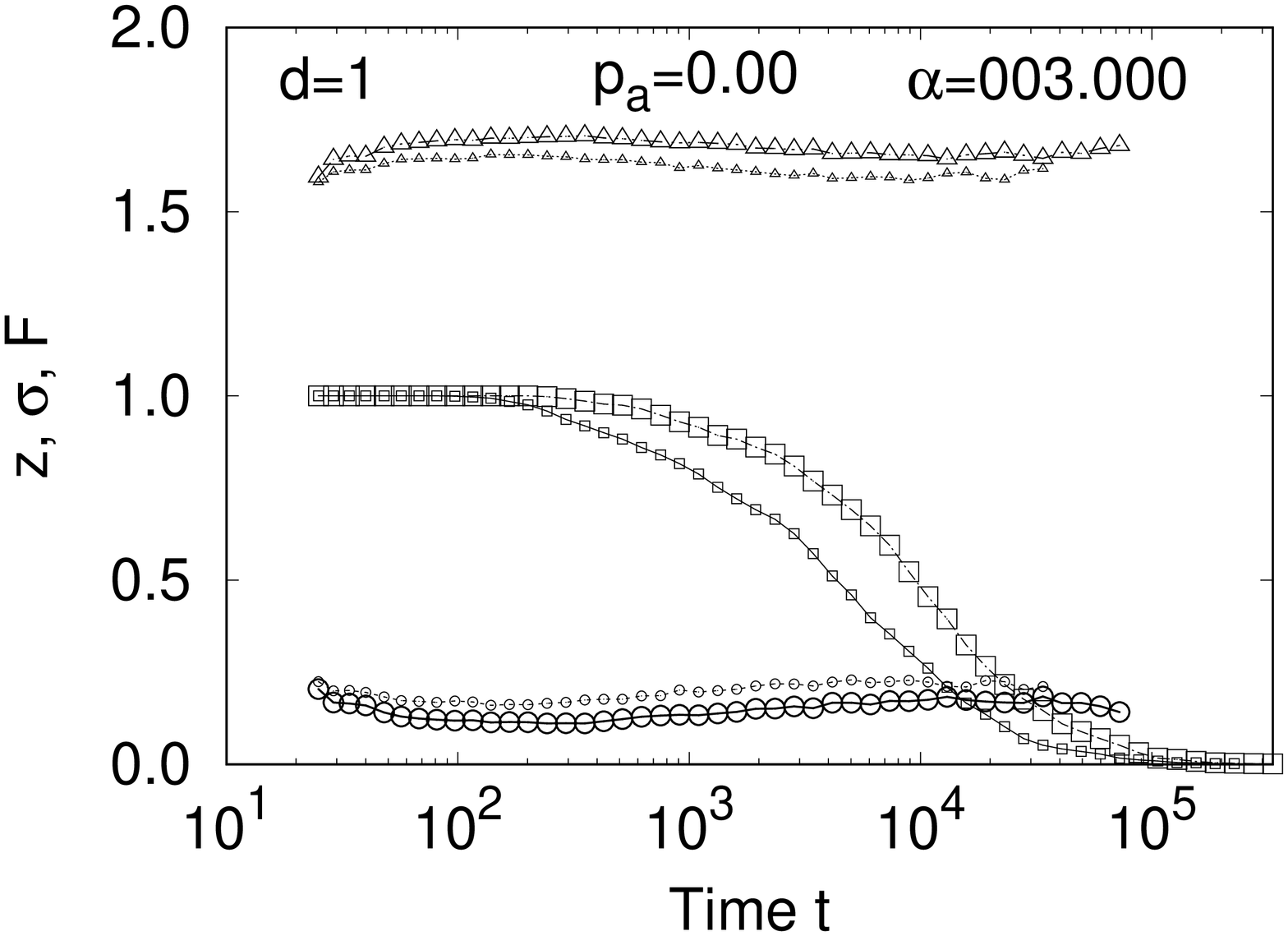}
  }
    \caption{Persistence (squares), Coherence $z$ (circles) and phase
      dispersion $\sigma$ (upwards triangles) vs time in
      one-dimensional networks with $L=200$ (small symbols) and
      $L=400$ (large symbols) for several values of the link power-law
      decay exponent $\alpha$.}
  \label{fig:Fzsvstime1D}
\end{figure}
\begin{figure}[h!]
    \centerline{
      \includegraphics[width=0.38\linewidth,angle=270]{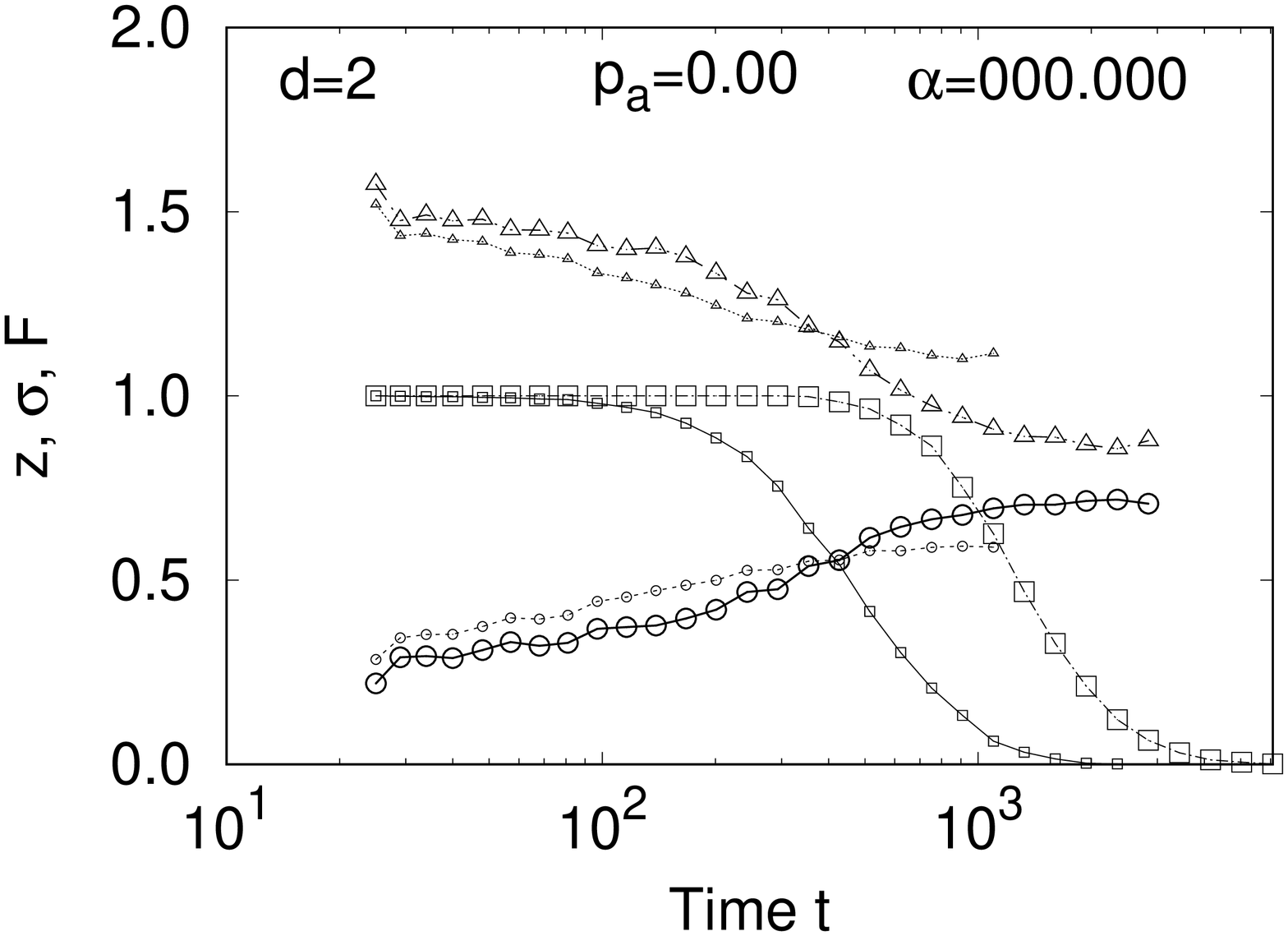}
      \includegraphics[width=0.38\linewidth,angle=270]{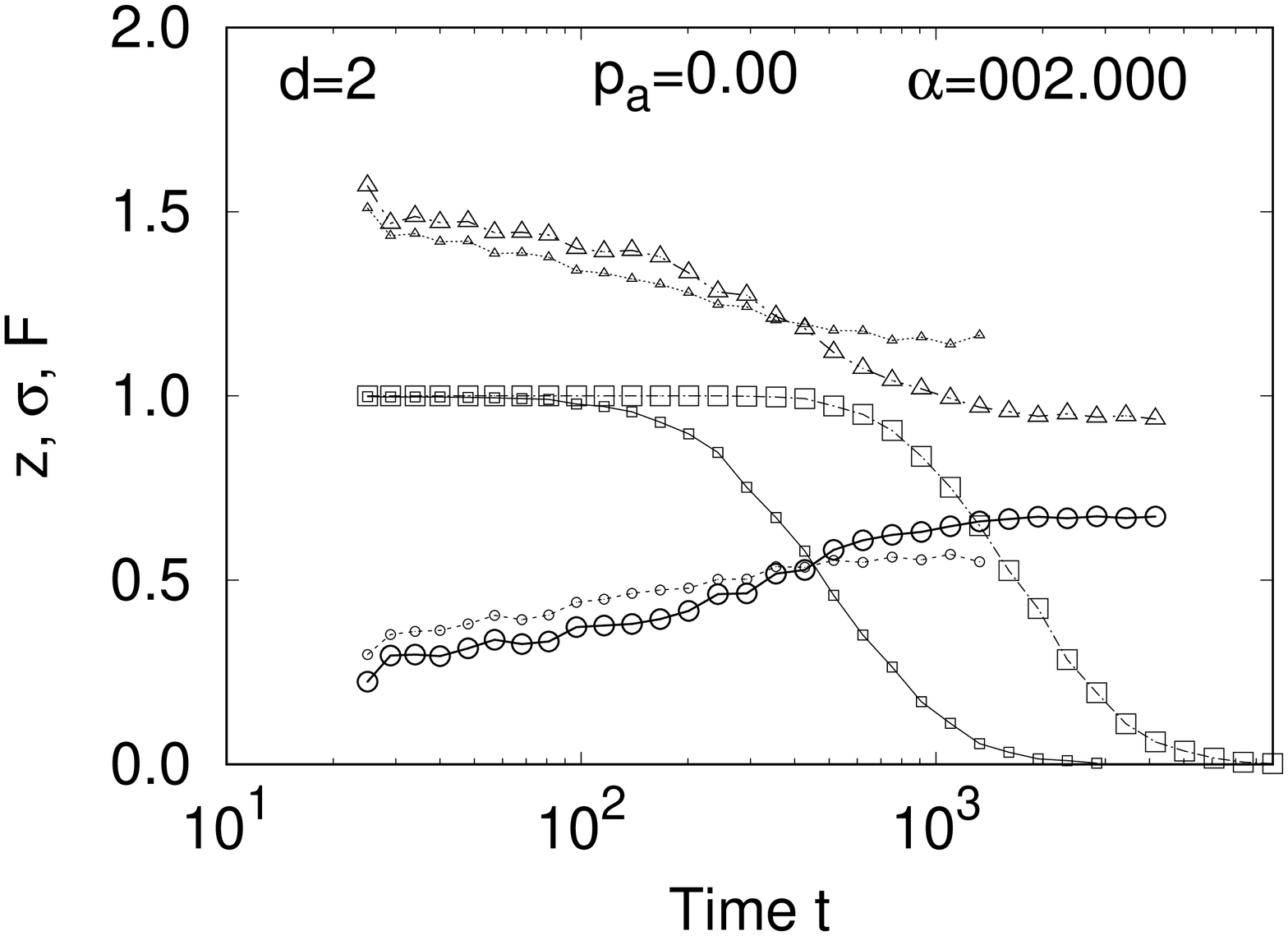}
    }
    \centerline{
      \includegraphics[width=0.38\linewidth,angle=270]{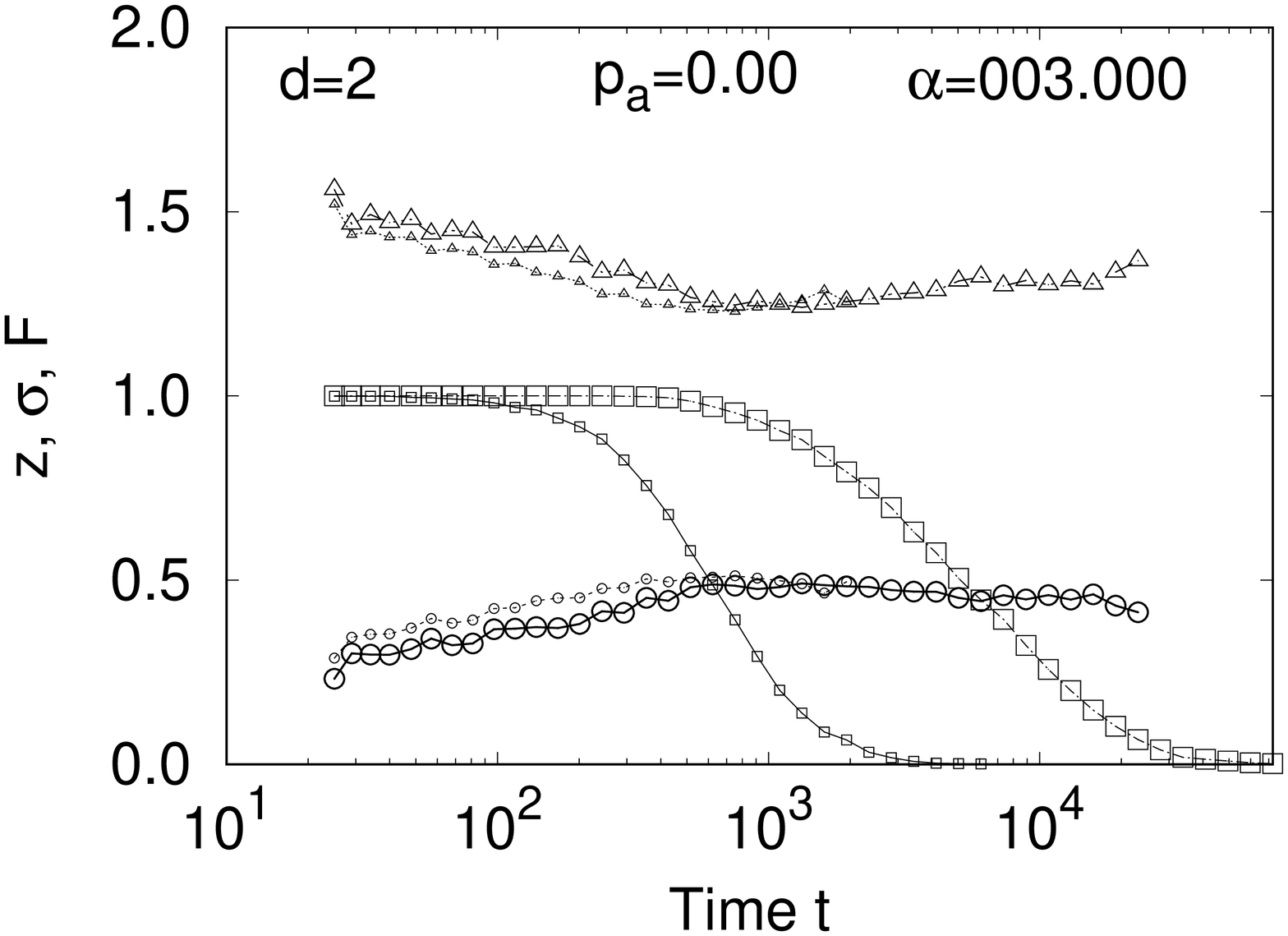}
      \includegraphics[width=0.38\linewidth,angle=270]{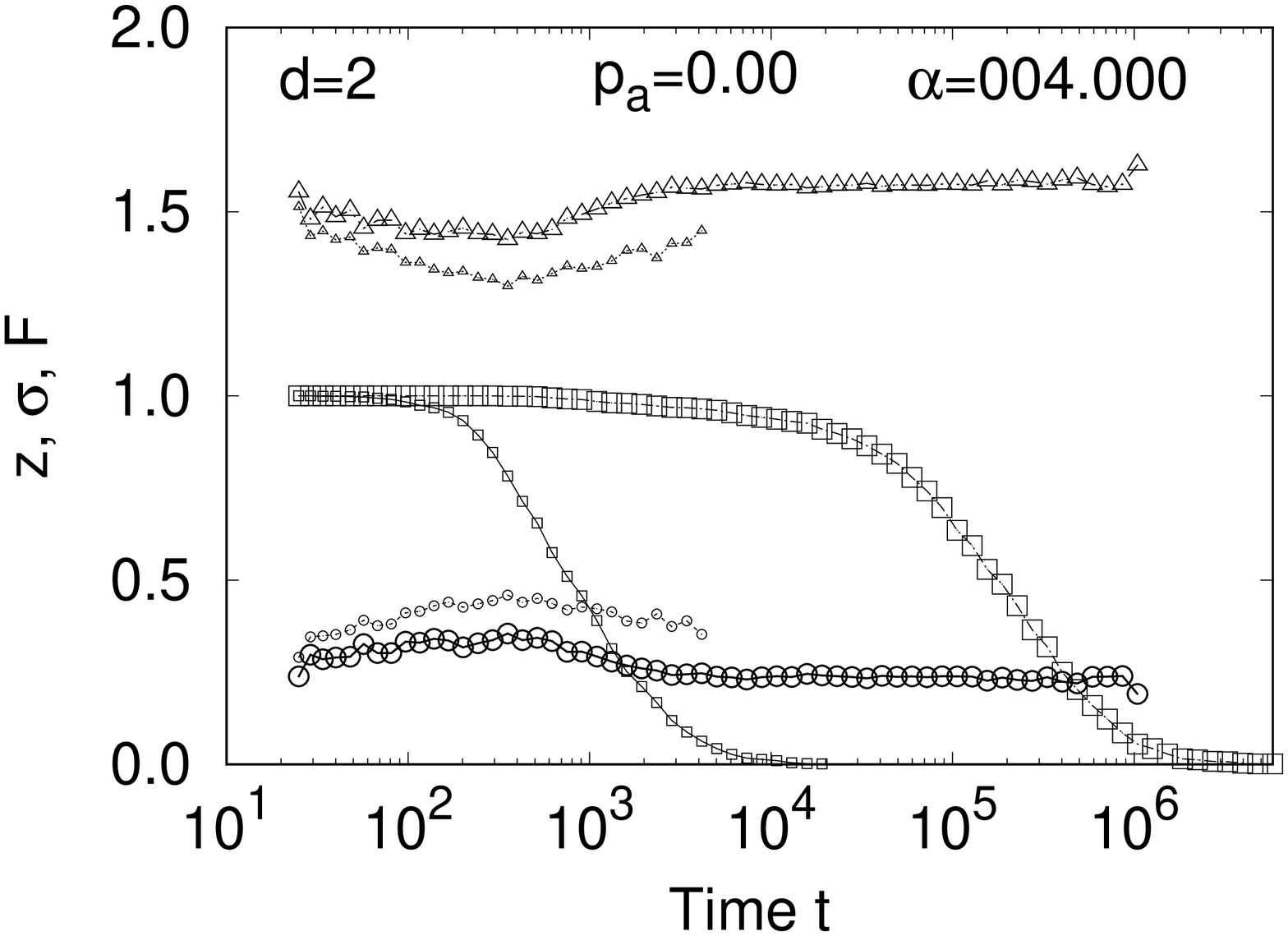}
    }
    \caption{Persistence (squares), Coherence $z$ (circles) and phase
      dispersion $\sigma$ (upwards triangles) vs time in
      two-dimensional networks with $L=10$ (small symbols) and $L=20$
      (large symbols) for several values of the link power-law decay
      exponent $\alpha$.}
  \label{fig:Fzsvstime2D}
\end{figure}

\begin{figure}[h!]
  \centerline{
    \includegraphics[width=0.38\linewidth,angle=270]{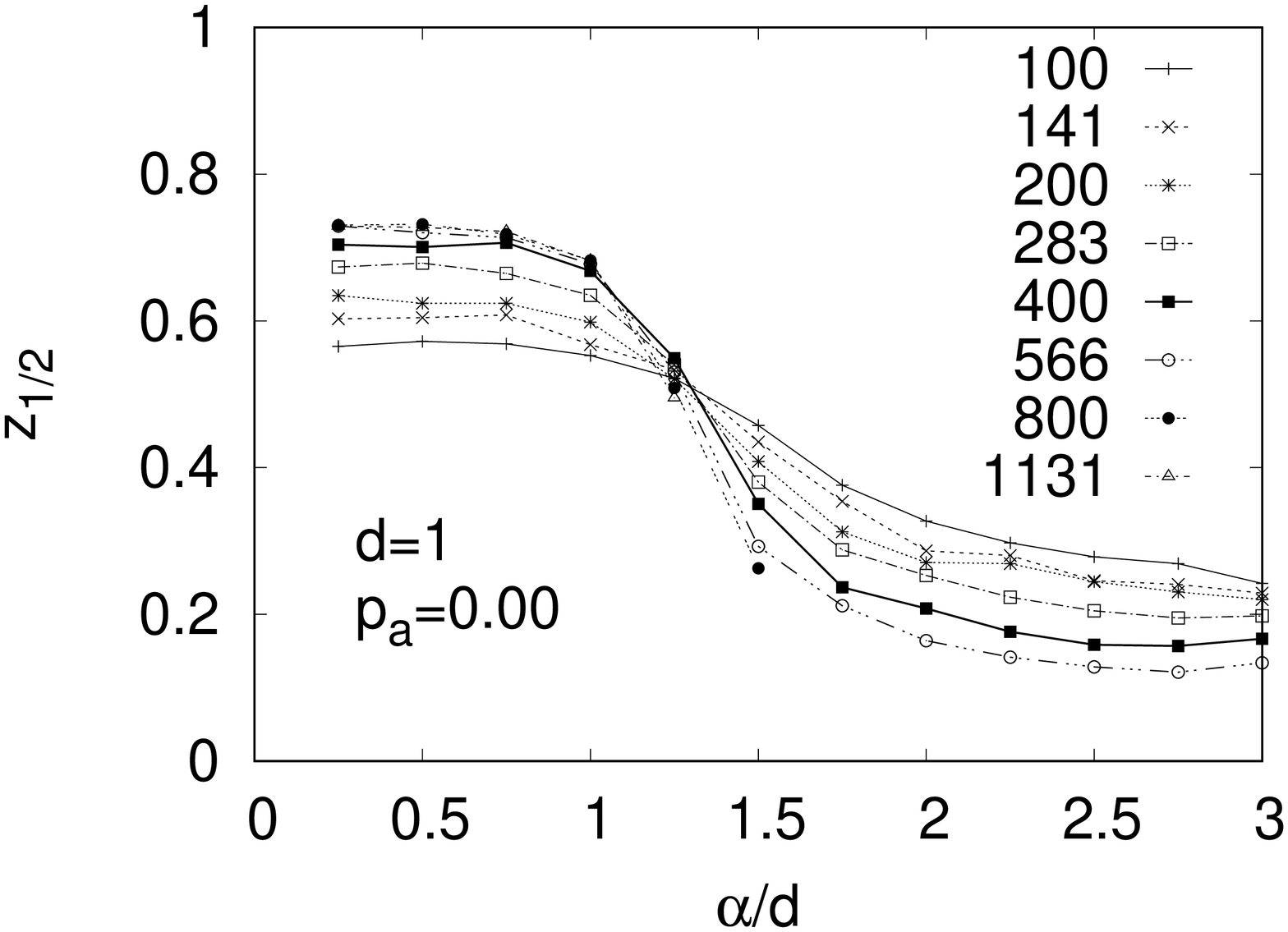}
    \includegraphics[width=0.38\linewidth,angle=270]{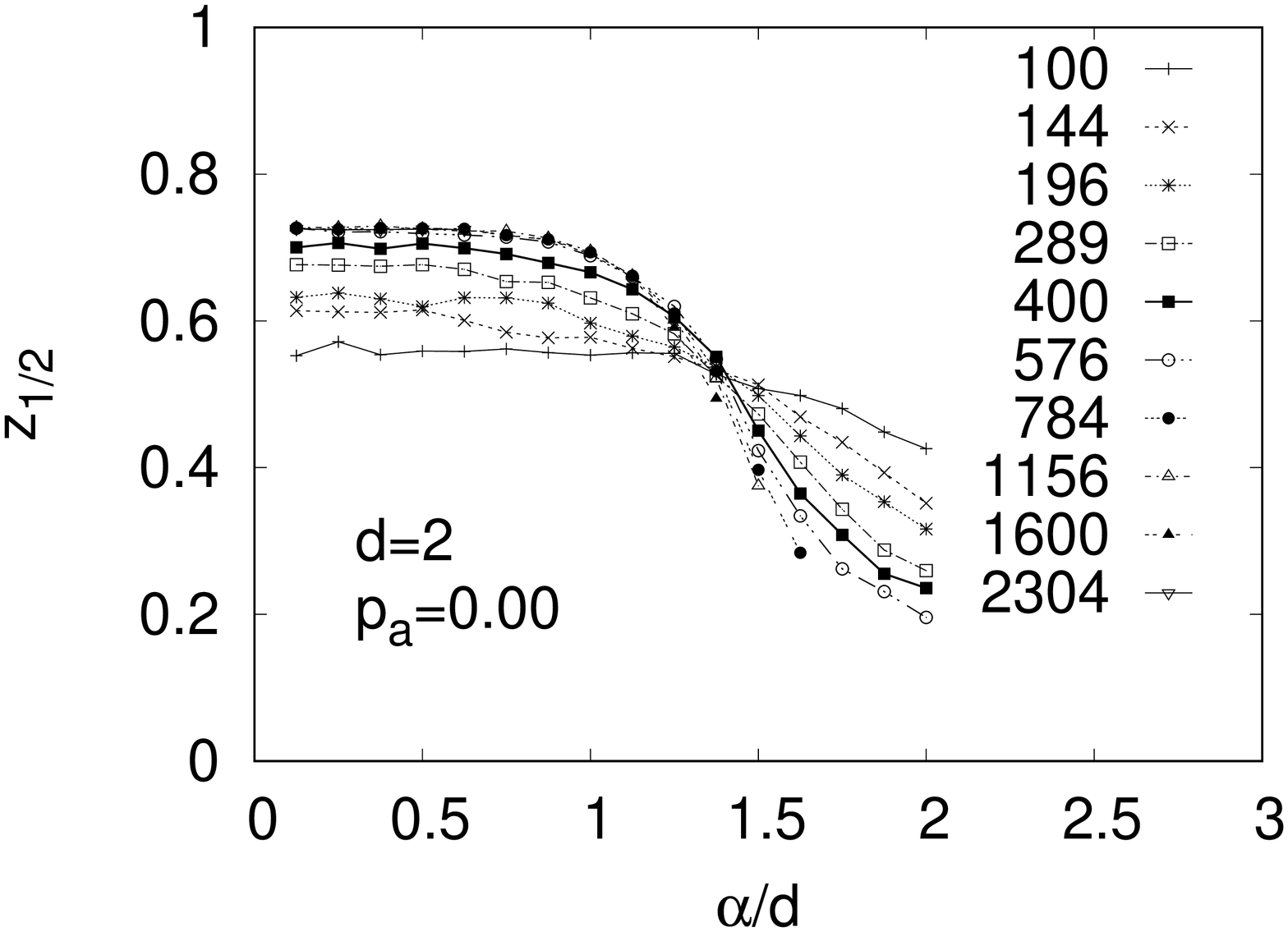}
  }
  \centerline{
    \includegraphics[width=0.38\linewidth,angle=270]{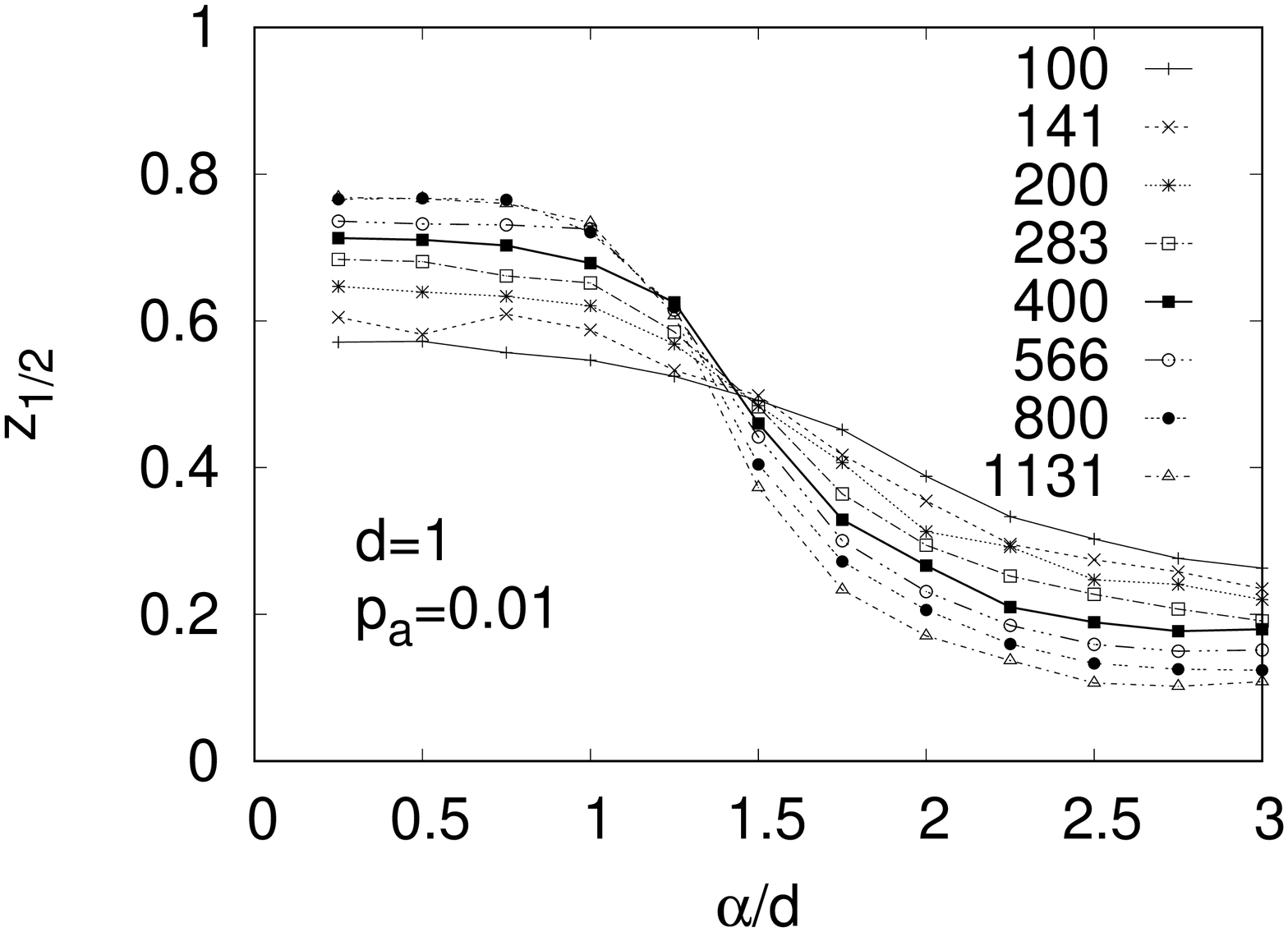}
    \includegraphics[width=0.38\linewidth,angle=270]{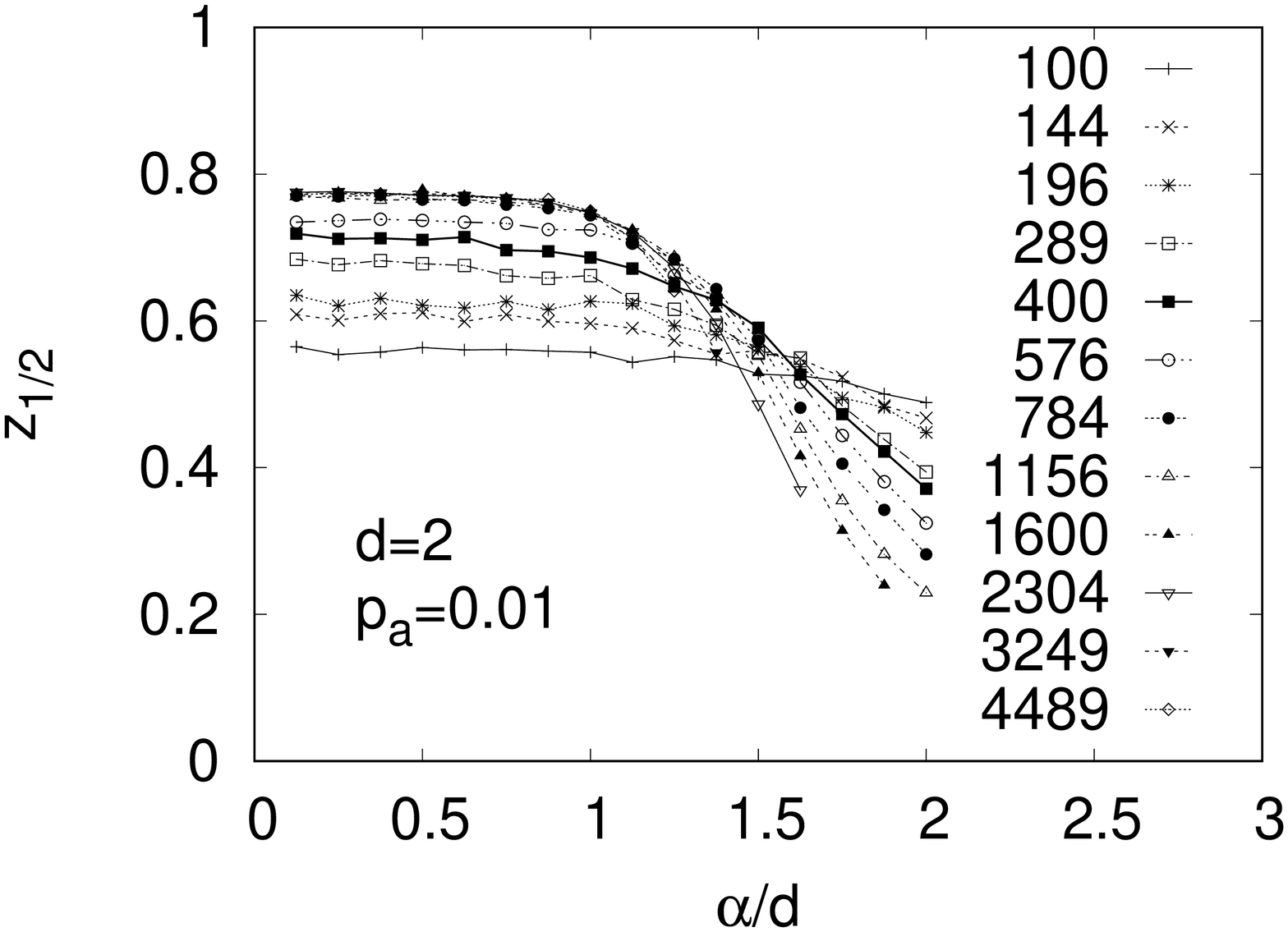}
  }
  \centerline{
    \includegraphics[width=0.38\linewidth,angle=270]{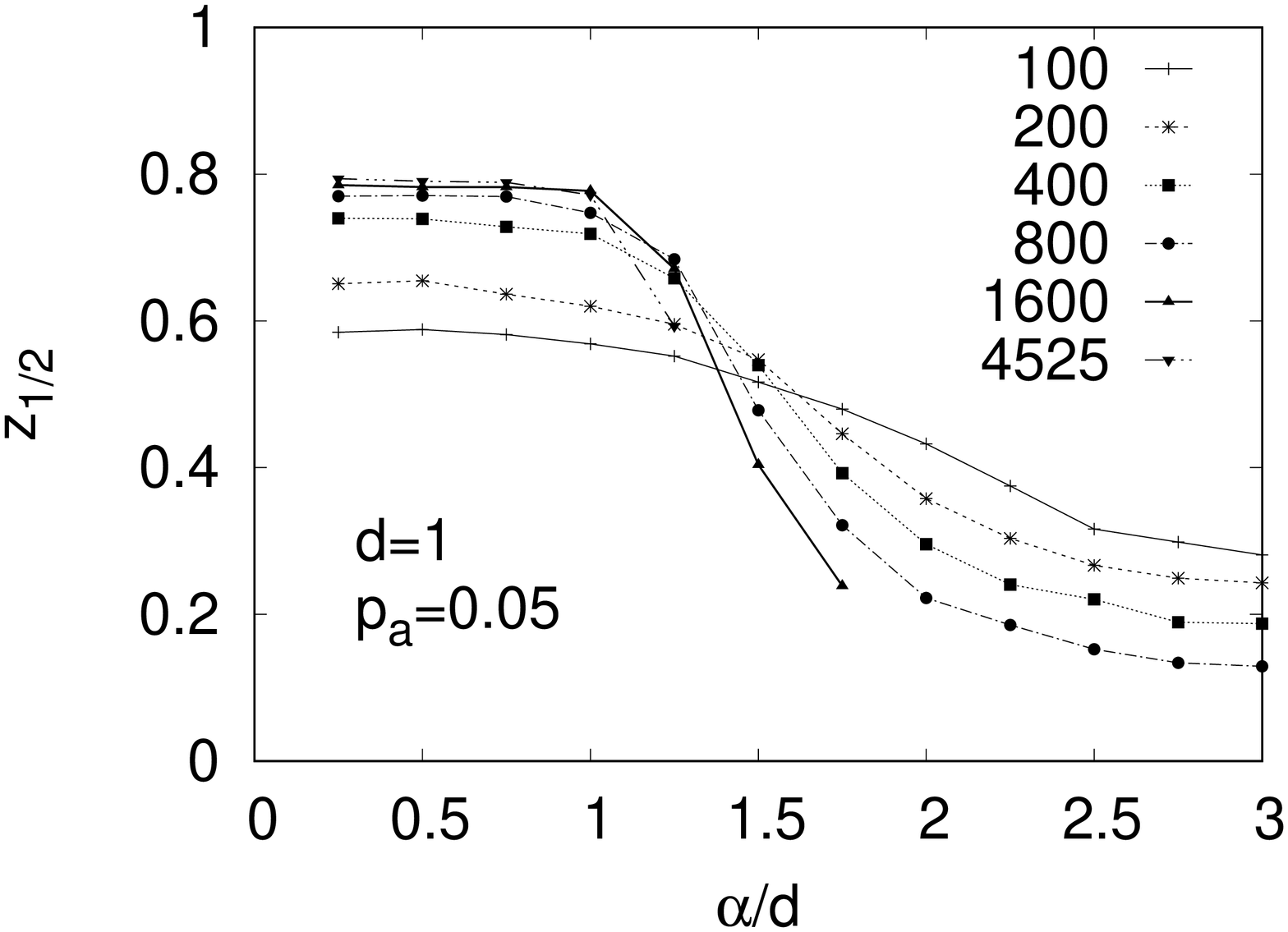}
    \includegraphics[width=0.38\linewidth,angle=270]{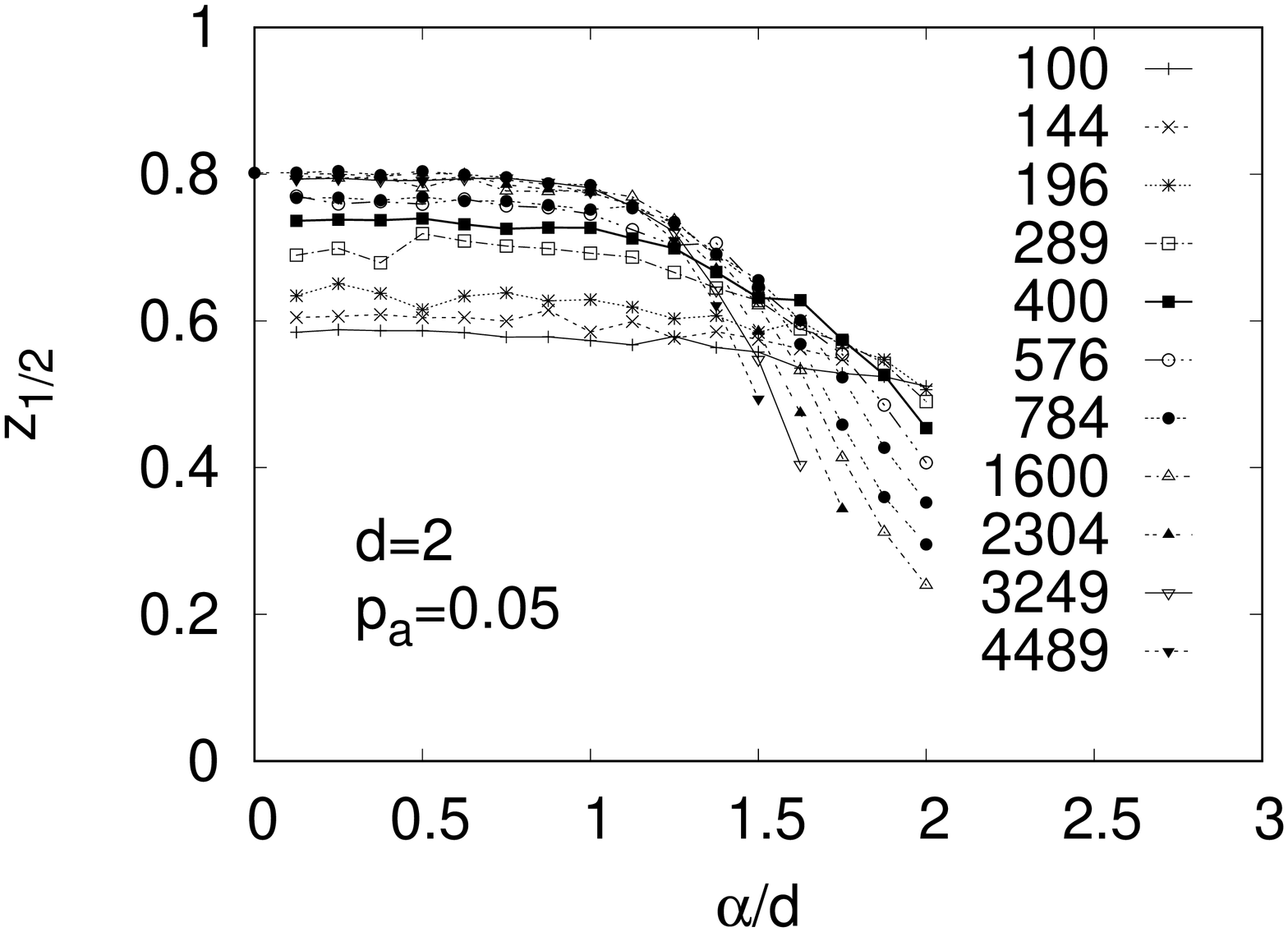}
  }
  \caption{Coherence $z$, averaged over 1000 networks, taken at
    halftime $t_{1/2}$ for extinction, vs link-length exponent
    $\alpha$, for several system sizes $N$ (labels of symbols in each
    plot) and for different amounts of link annealing $p_a$. Plots on
    the left are for one-dimensional networks, those on the right for
    two-dimensional ones. These results suggest that a synchronized
    second stage exists only for $\alpha < d$ in both one and two
    dimensions. Symbol labels indicate the number $N$ of sites in the
    system. }
  \label{fig:coherence.at.halftime.vs.alpha}
\end{figure}

Let us begin by discussing the basic observable features of the
synchronization process in these networks. As mentioned previously, we
have chosen a value for the link infection probability that is large
enough to ensure synchronization-induced extinctions. Therefore, we
expect that spontaneous extinctions will not be relevant.
\\ \Figs{fig:Fzsvstime1D} and \ref{fig:Fzsvstime2D} show the coherence
$z$, the phase variance $\sigma$, and the persistence $F$, resulting
from averages over $10^3$ realizations in $d=1$ and $2$ dimensions,
for chosen values of the link exponent $\alpha$, on static ($p_{a}=0$)
networks. Networks with link-annealing ($p_{a}>0$) produce similar
results, but on shorter timescales. At the beginning of a simulation,
$z$ is low and $\sigma$ is large, because node phases are initially
distributed at random over the active period. When $\alpha \leq d$,
two dynamical stages are clearly distinguishable: The first one is the
stage of \emph{Initial Synchronization}.  During this stage, the
coherence $z$ increases and the phase dispersion $\sigma$ decreases,
until both stabilize after approximately $10^2$ to $10^3$ steps. No
extinctions occur during this initial organization period, i.e.~the
persistence $F(t)$ stays exactly at one. Next comes the stage of
\emph{Sustained coherence}. During this second stage, the average
coherence $z$ shows a constant value, and so does
$\sigma$. Extinctions start to occur in this stage, as evidenced by a
decreasing value of the persistence $F(t)$.  Notice, however, that for
$\alpha > d$, synchronization during stage two diminishes with
increasing $\alpha$. In this case, after a short-lived synchronization
period, $z$ decreases and $\sigma$ increases again in time
(\Figs{fig:Fzsvstime1D} and \ref{fig:Fzsvstime2D}). Synchronization,
as measured by $z$, becomes weaker for larger systems
(\Fig{fig:coherence.at.halftime.vs.alpha}), suggesting that no
synchronized stage exists in the thermodynamic limit for large
$\alpha$. Networks with large $\alpha$ are topologically
$d$-dimensional because all links are short-ranged.
In order to further explore the $\alpha$-dependence of
synchronizability in these networks, we measured $z_{1/2}$, the
average coherence at the median time $t_{1/2}$ for extinction, (when
half the samples still survive). These results are displayed in
\Fig{fig:coherence.at.halftime.vs.alpha}. They suggest that
synchronization only happens, in the thermodynamic limit, for $\alpha
\leq d$, in both one and two dimensions, although more extensive work
would be needed in order to determine the critical value of $\alpha$
precisely. Furthermore, the value of the annealing parameter $p_{a}$
does not seem to modify these static results significantly, although
annealing does speed up the dynamics of the synchronization process
(see later).

\subsection{Late-stage extreme synchronization}
\label{latestage}
\begin{figure}[h!]
    \centerline{
      \includegraphics[width=0.38\linewidth,angle=270]{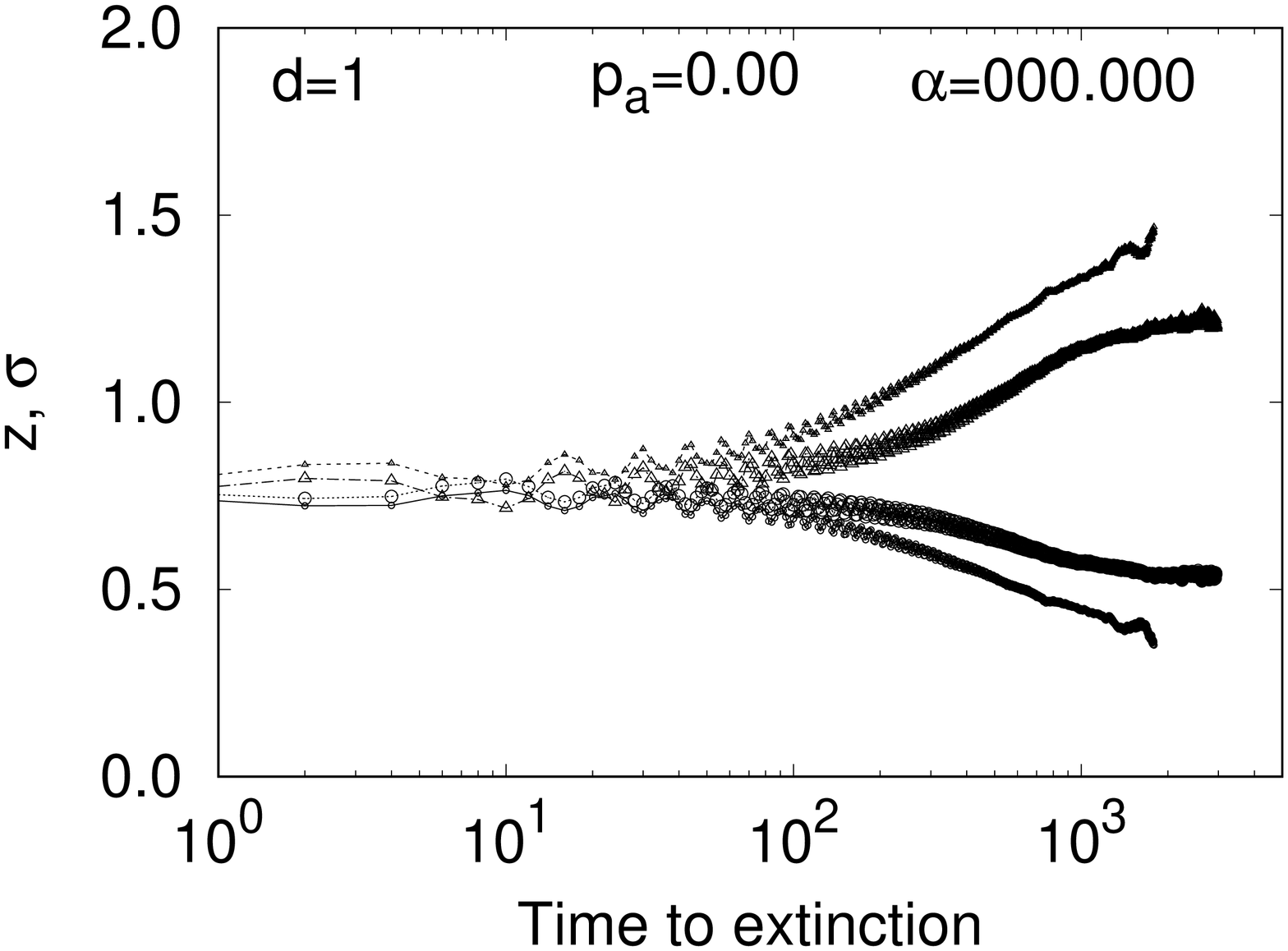}
      \includegraphics[width=0.38\linewidth,angle=270]{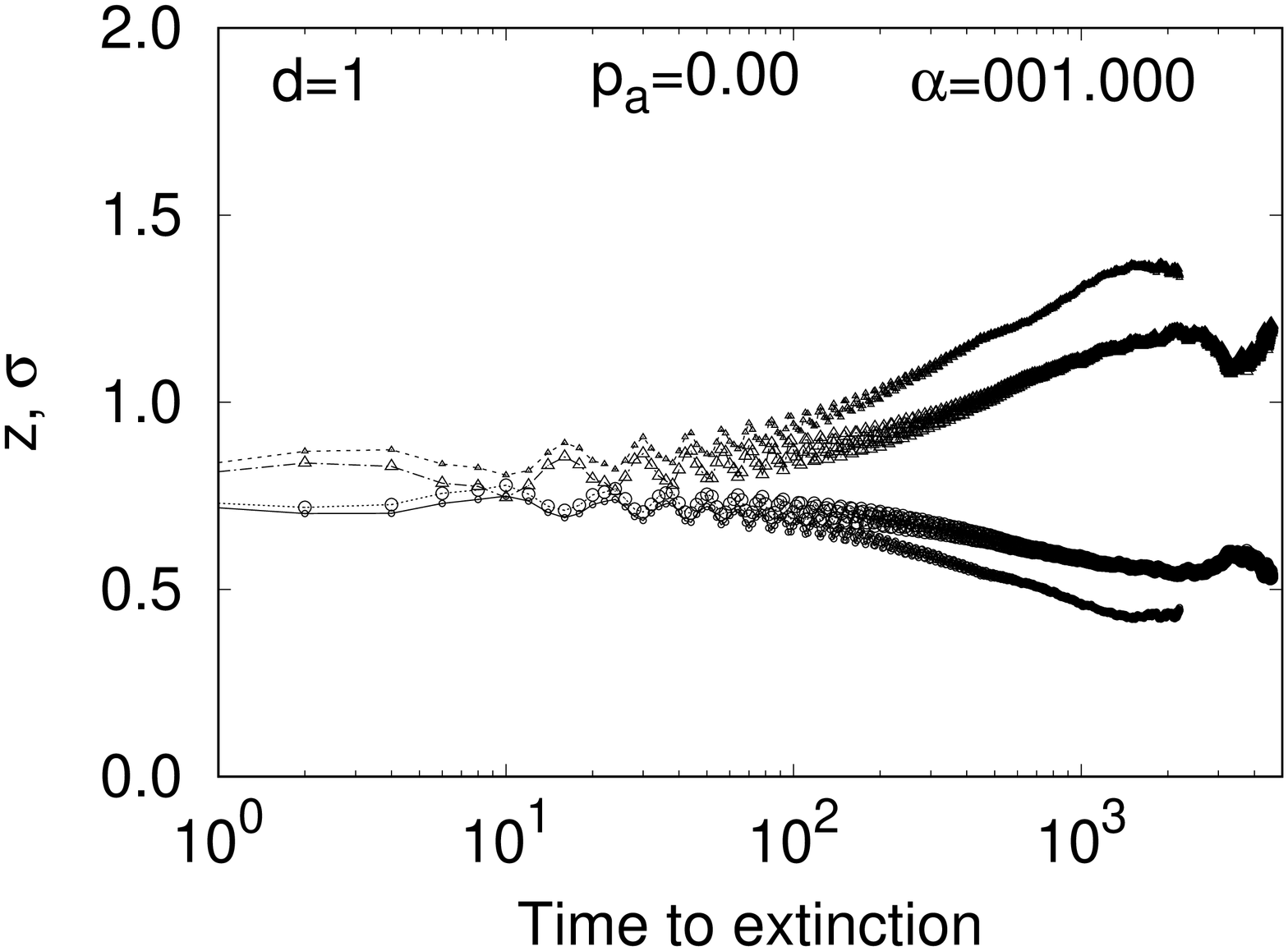}
    }
    \centerline{
      \includegraphics[width=0.38\linewidth,angle=270]{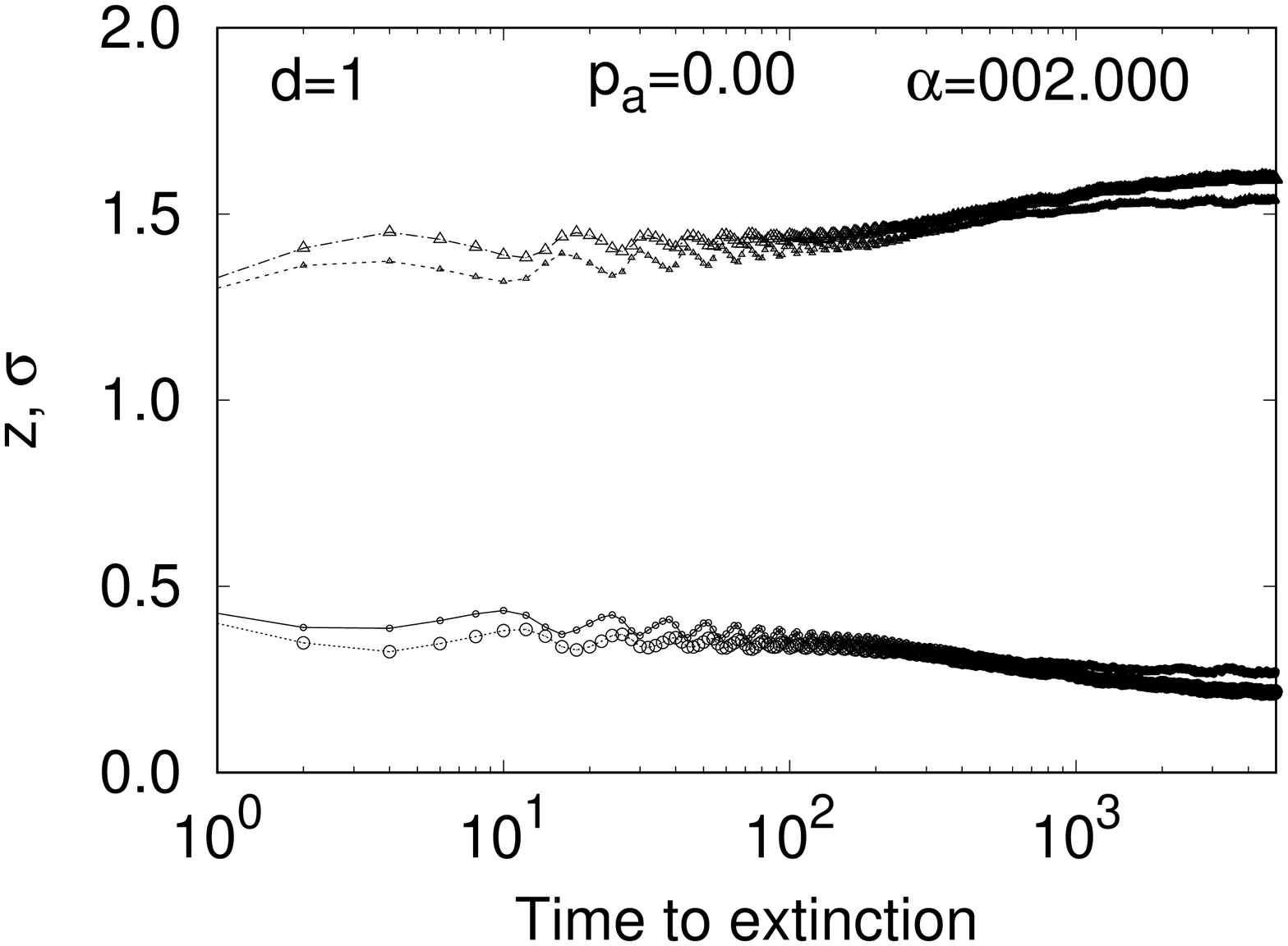}
      \includegraphics[width=0.38\linewidth,angle=270]{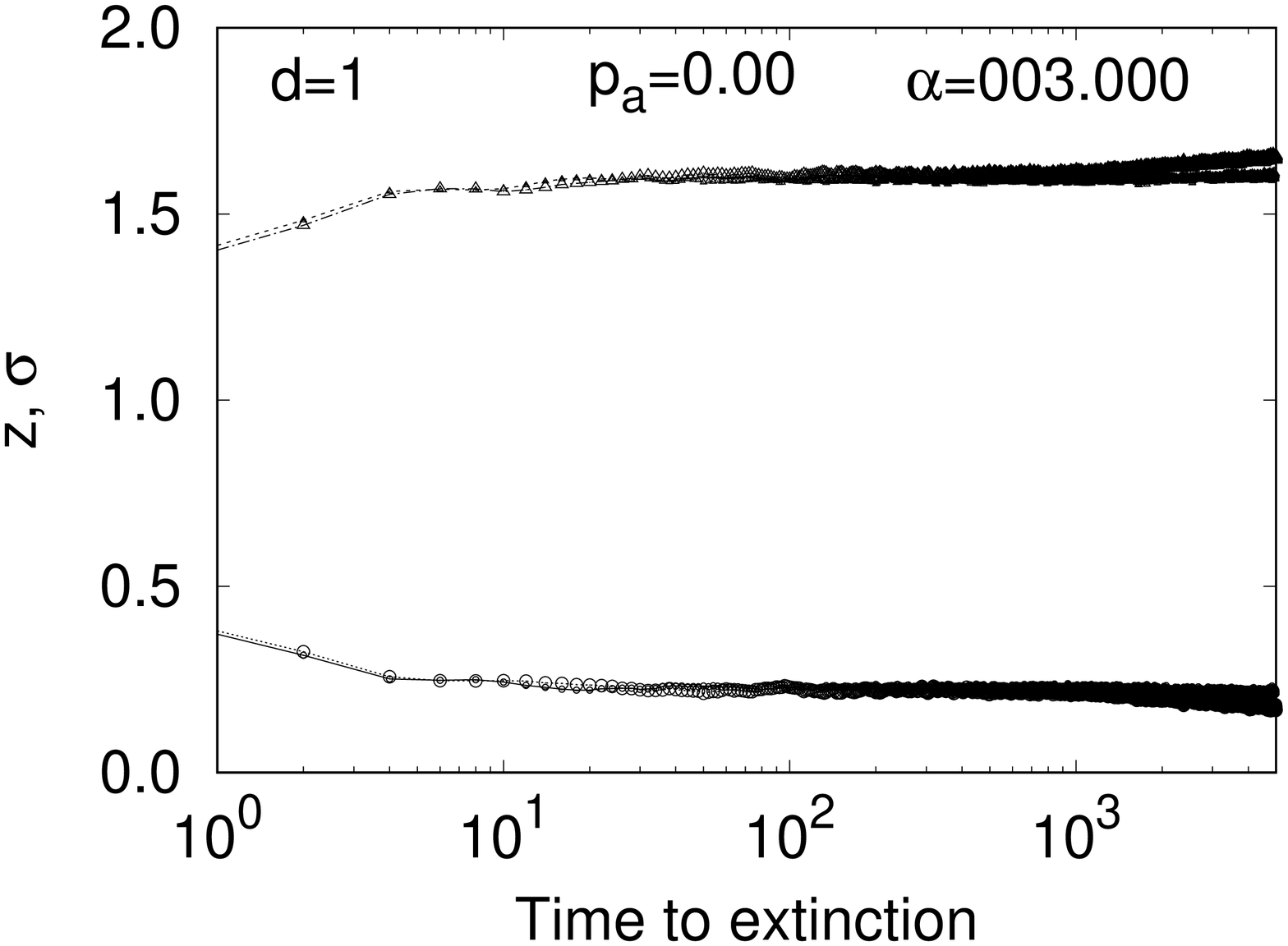}
    }
    \caption{Coherence $z$ (circles) and phase dispersion $\sigma$
      (upwards triangles) vs time-to-extinction in one-dimensional
      networks with $L=200$ (small symbols) and $L=400$ (large
      symbols) for several values of the link power-law decay exponent
      $\alpha$.}
  \label{fig:Fzsvstime1D.R}
\end{figure}

\begin{figure}[h!]
  \centerline{
    \includegraphics[width=0.38\linewidth,angle=270]{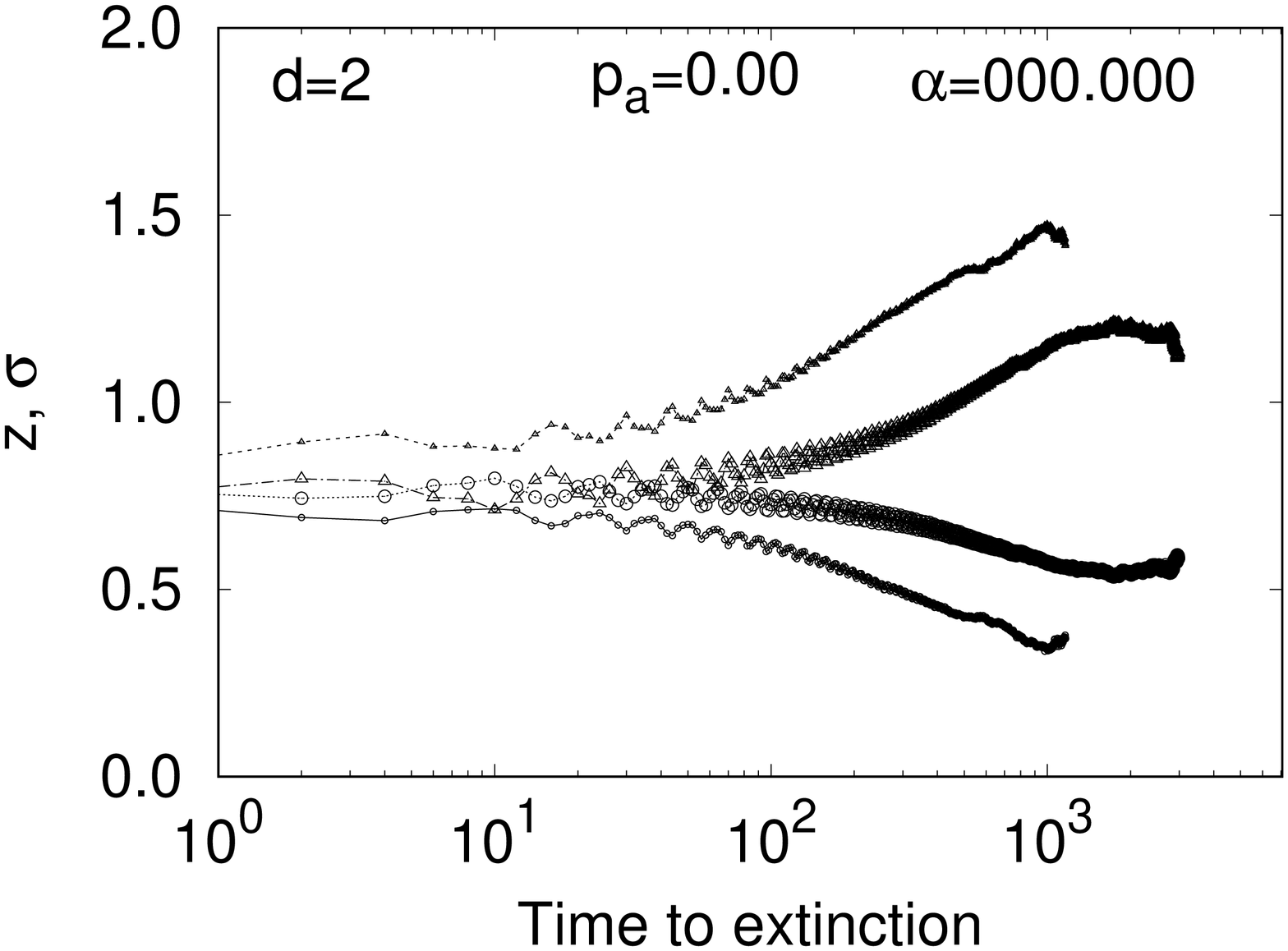}
    \includegraphics[width=0.38\linewidth,angle=270]{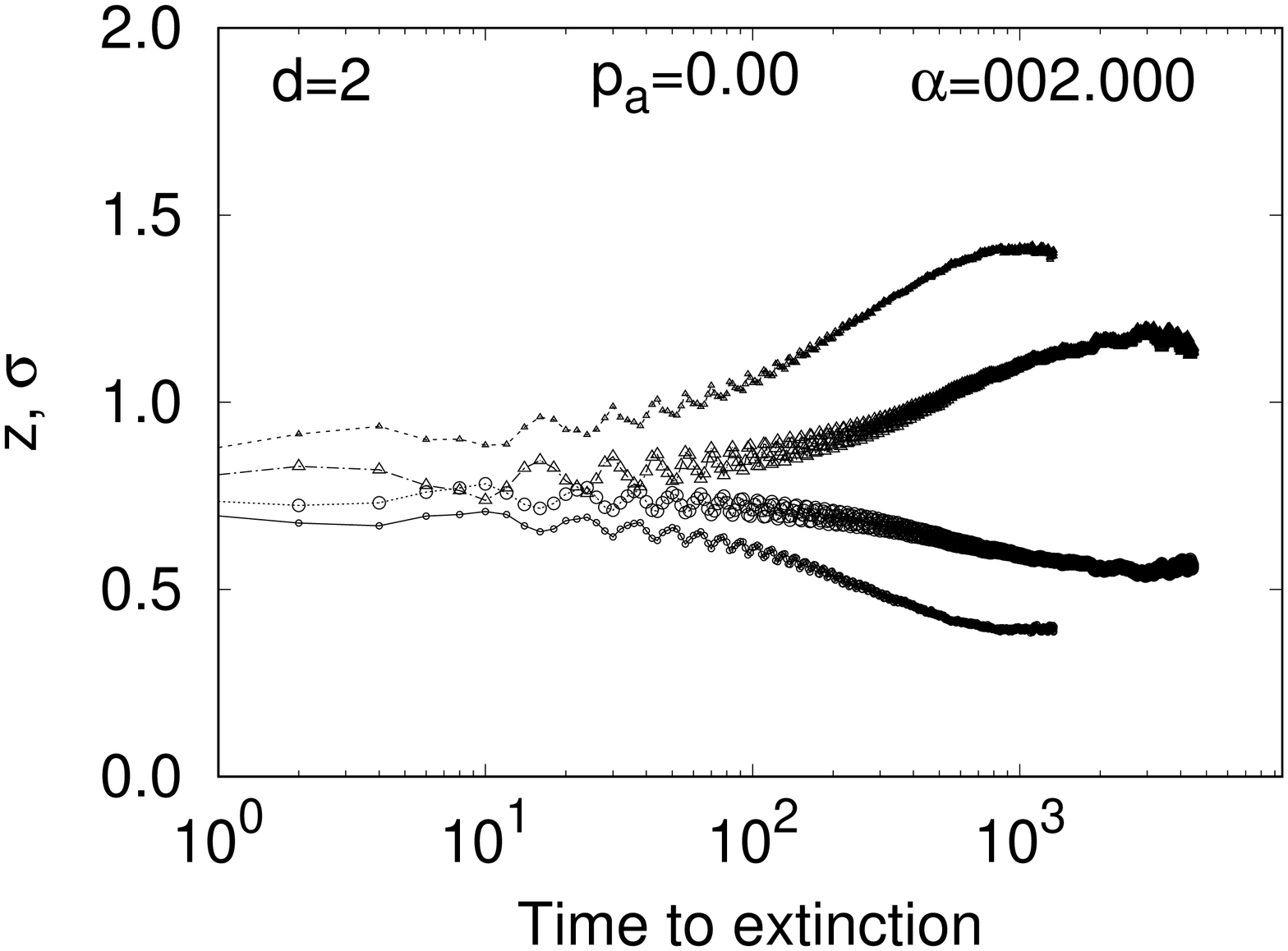}
  }
  \centerline{
    \includegraphics[width=0.38\linewidth,angle=270]{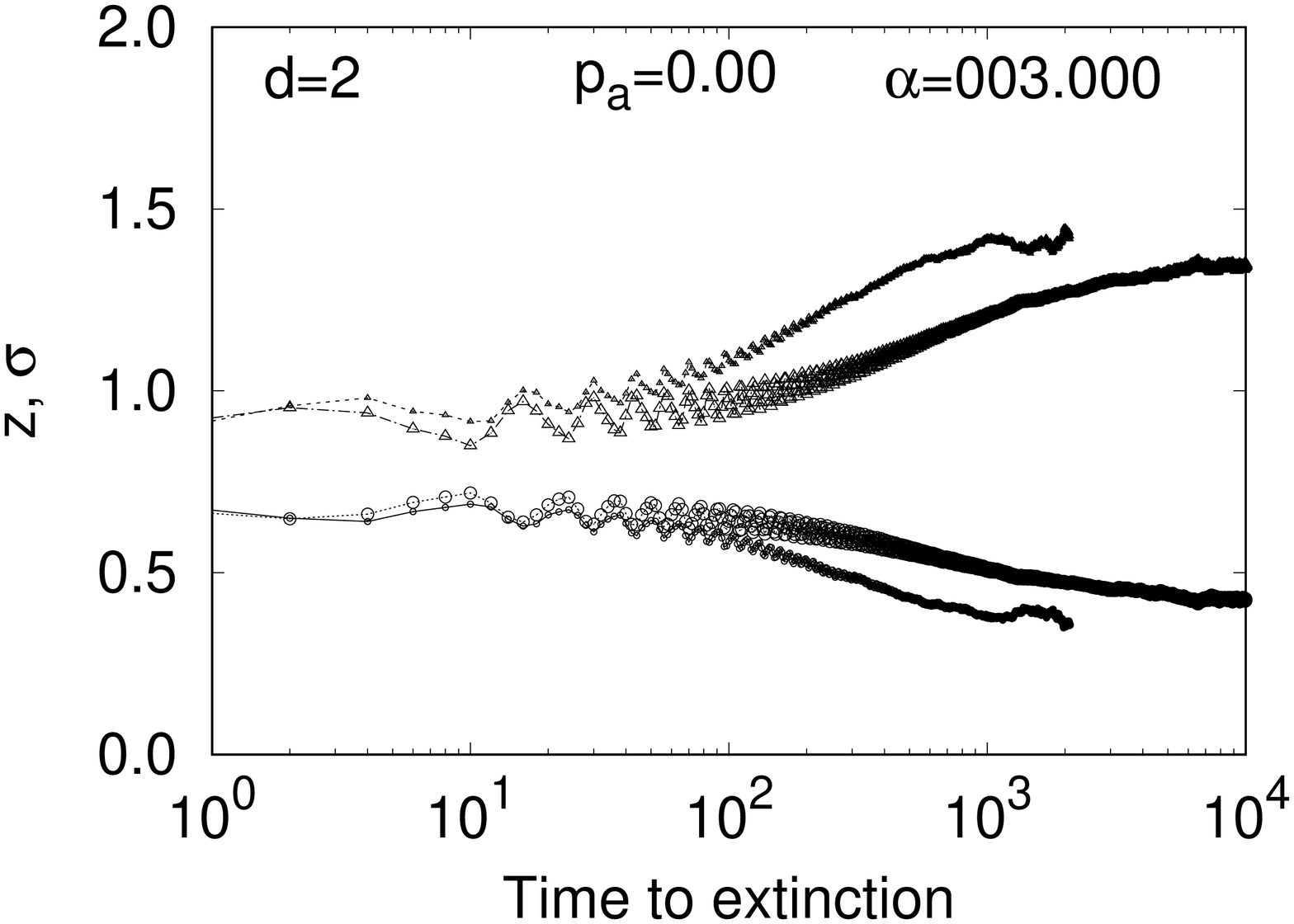}
    \includegraphics[width=0.38\linewidth,angle=270]{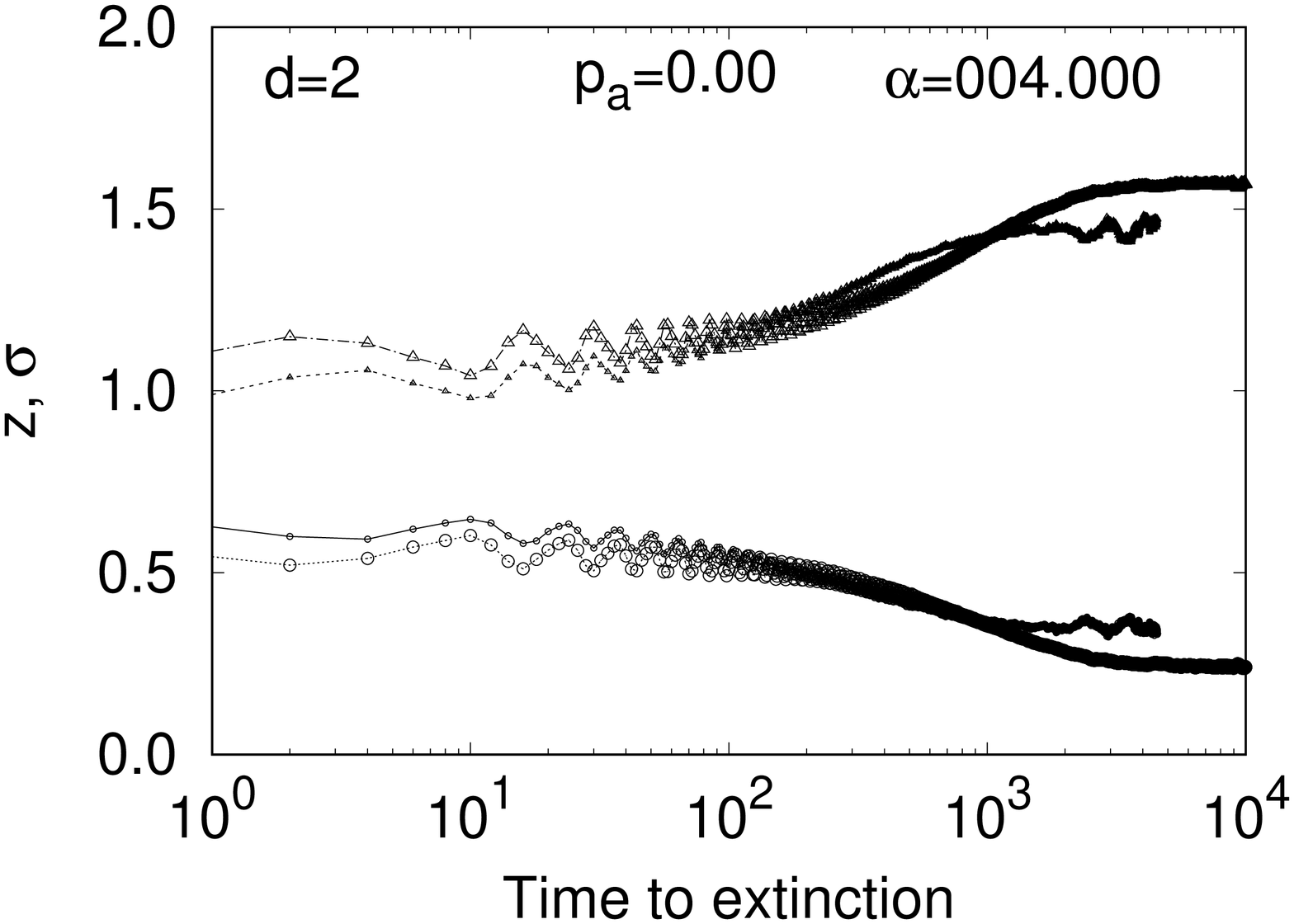}
  }
  \caption{Coherence $z$ (circles) and phase dispersion $\sigma$
    (upwards triangles) vs time-to-extinction in two-dimensional
    networks with $L=10$ (small symbols) and $L=20$ (large symbols)
    for several values of the link power-law decay exponent $\alpha$.}
  \label{fig:Fzsvstime2D.R}
\end{figure}

The data displayed in \Figs{fig:Fzsvstime1D} and \ref{fig:Fzsvstime2D}
would seem to suggest that extinctions happen while the system is in a
state of partial synchronization (in stage two). Actually, however, this
is not true. Extinction events are in fact invariably preceded by a
short burst of extreme synchronization. In this last stage, which
inevitably leads to extinction, the coherence $z$ increases, and the
phase dispersion $\sigma^2$ decreases rather abruptly, within roughly
$10^3$ timesteps (See \Figs{fig:Fzsvstime1D.R} and
\ref{fig:Fzsvstime2D.R}).  The above mentioned late-stage increases in
coherence are not seen in \Figs{fig:Fzsvstime1D} and
\ref{fig:Fzsvstime2D}, because the distribution of extinction times is
so broad that these short-lived bursts of extreme synchronization do
not contribute to the sample-averages.

In order to clearly reveal the true dynamical behavior of the system
immediately before extinction, we recorded, for each simulation,
results from the last $10^4$ timesteps (using a circular array). These
data were subsequently averaged in such a way that extinction times
coincide for all simulations, and taking averages at equal values of
the \emph{time-to-extinction} $t_{te}$.  These measurements in
``reverse time'' are displayed in \Figs{fig:Fzsvstime1D.R} and
\ref{fig:Fzsvstime2D.R} versus $t_{te}$. They show that the coherence
$z$ increases steeply, and the phase dispersion $\sigma$ decreases,
right before extinction.  Therefore, at a random time and while on
stage two, the system starts to synchronize even more, and does so
rather rapidly. Approximately $10^2$ to $10^3$ steps later, extinction
of the dynamics invariably happens.  Therefore, some random event (see
later for a discussion) seems to occur during stage two, after which the
system inevitably undergoes extreme synchronization and becomes
extinct.  Notice that different realizations of the experiment are
synchronized \emph{with each other}, when considered at equal values
of $t_{te}$, as evidenced by the oscillations seen in
\Figs{fig:Fzsvstime1D.R} and \ref{fig:Fzsvstime2D.R}.  This implies
that all extinctions occur at the same point during the oscillatory
phase of the system, and also that the final steps of the dynamics are
similar, and occurr in phase with each other, for all network
realizations.  We can therefore conclude that there is in fact a third
dynamical stage, with an approximate duration of $10^2-10^3$ timesteps
for most of the cases analyzed here, during which the system undergoes
extreme synchronization, and which inevitably leads to the extinction
of the dynamics.  

We furthermore notice that stages one and three are relatively short,
while stage two is by far the longest (for large systems), with an
average duration that increases very fast with system size. The time
that a system spends in stage two is a stochastic variable with a
distribution that broadens with increasing system size.  (see
\Sec{exp-dist-t2}).  The permanence time of the system in stages one
and three, on the other hand, is almost deterministic. Let us mention
that the initial ``resilience time'' $\tau_0$ during which the
persistence $F(t)$ stays at one, corresponds to the added durations of
stages one and three, which is the minimum amount of time after which
an extinction can happen. In this work we will not discuss how to
separately measure the length of stages one and three.
\begin{figure}[h!]
\centerline{
  \includegraphics[width=0.50\linewidth]{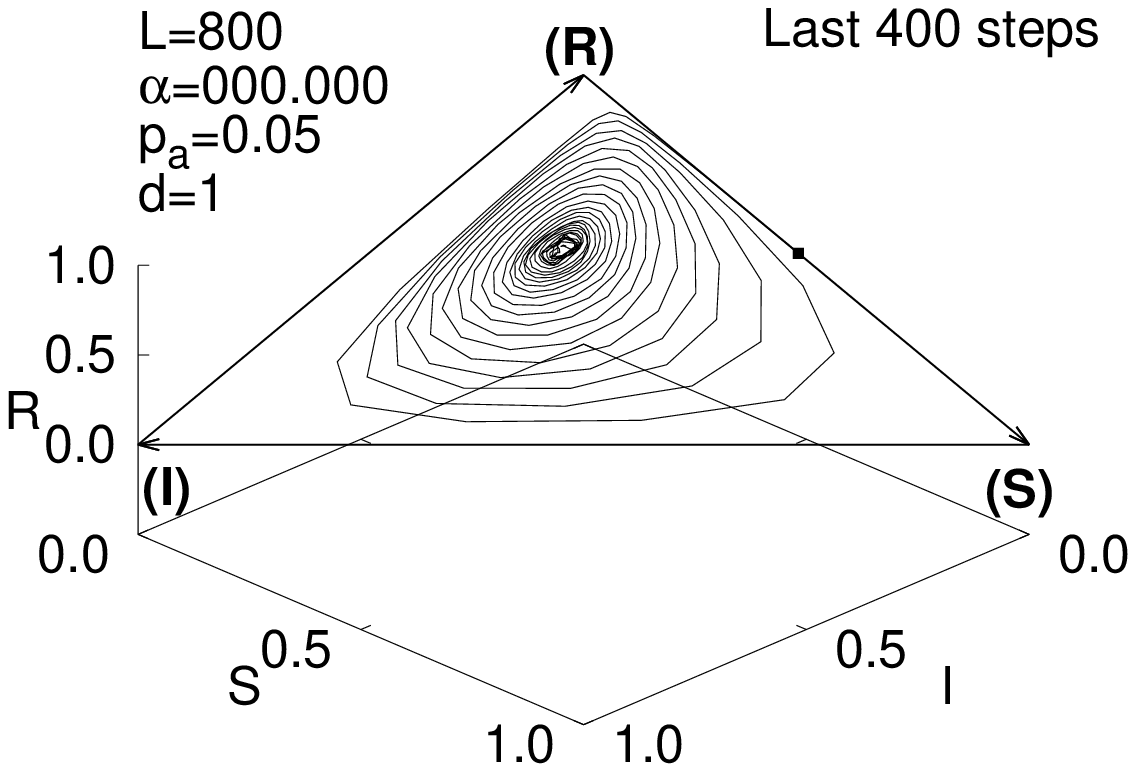}
  \includegraphics[width=0.50\linewidth]{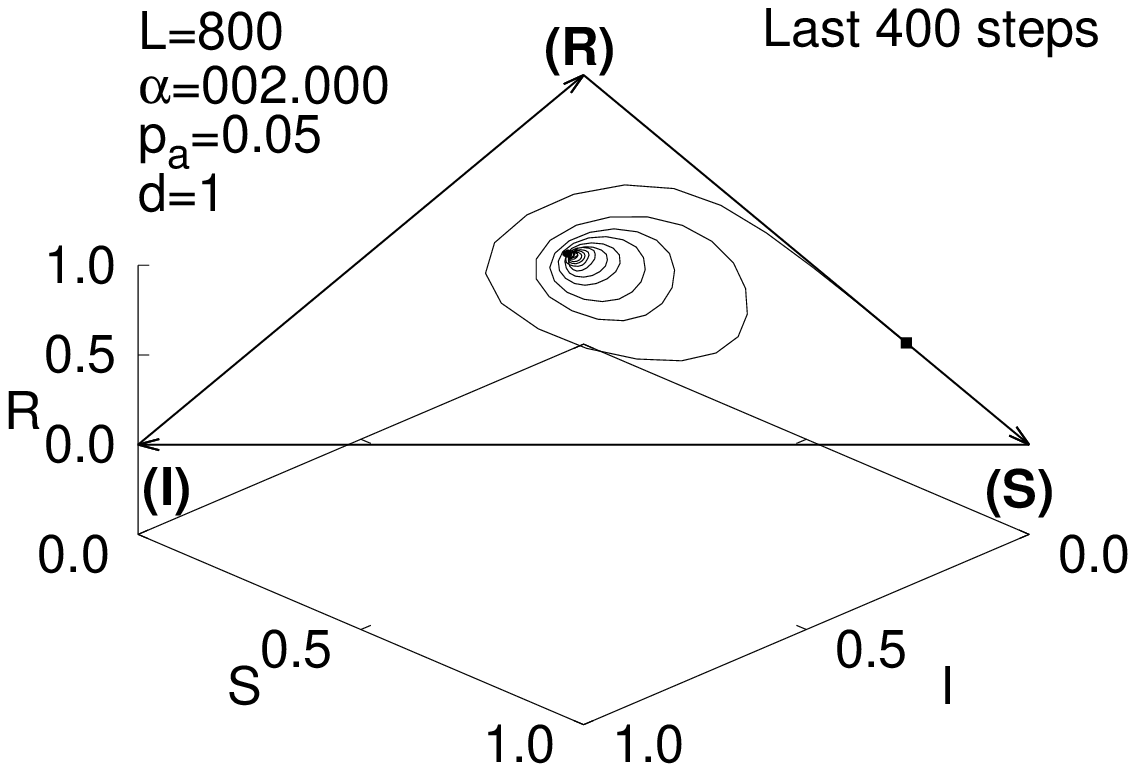}
}\centerline{
  \includegraphics[width=0.50\linewidth]{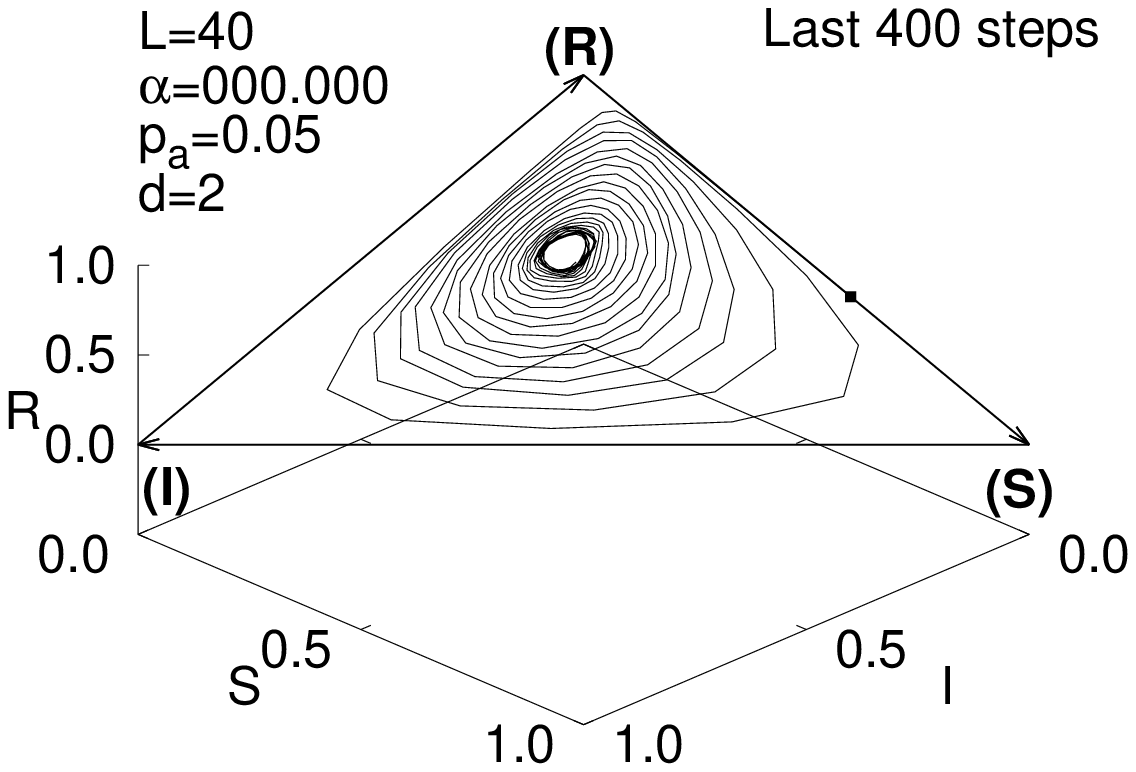}
  \includegraphics[width=0.50\linewidth]{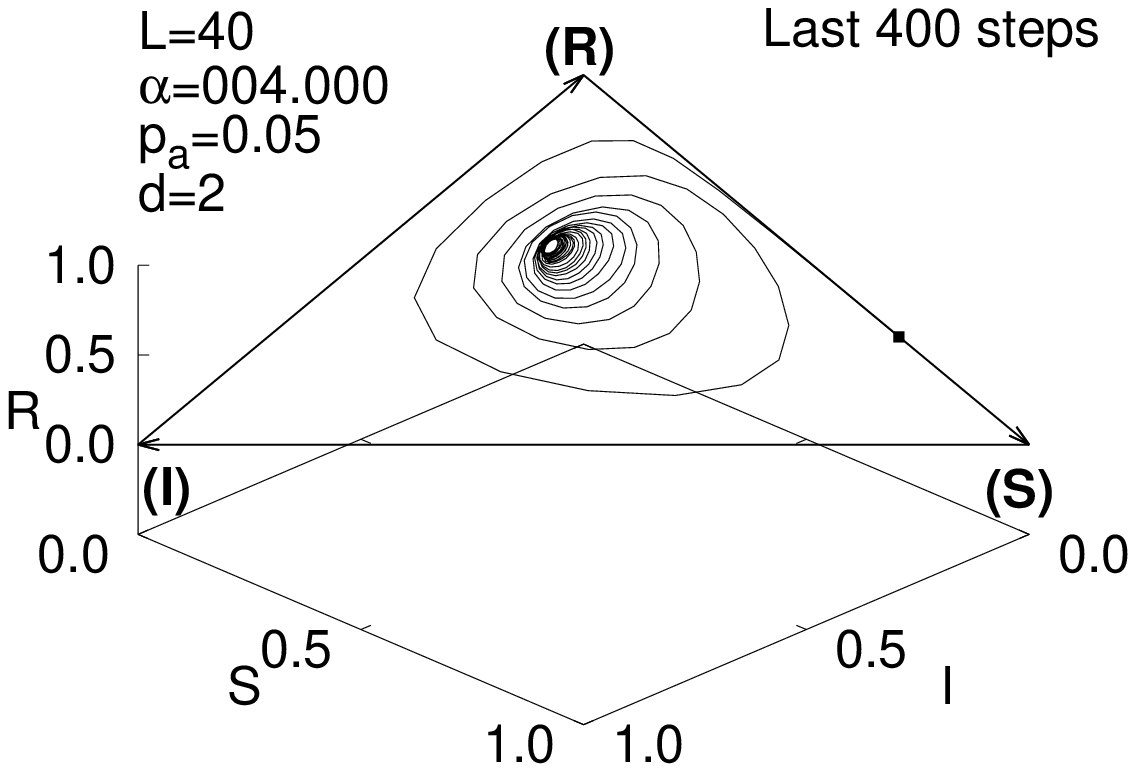}
}
\caption{ Phase-space location of the triad $\triad=(R,S,I)$,
  averaged over 1000 networks at equal time-to-extinction, in $d=1$
  (top) with $L = 800$, and $d=2$ (bottom) with $L=40$. Plots on the
  left are for $\alpha/d=0$, those on the right for $\alpha/d=2$.  In
  all cases the annealing rate per link is $p_{a} = 0.05$, and only
  the last 400 steps are shown. }
  \label{phase.space.rev}
\end{figure}

\Fig{phase.space.rev} shows the last 400 steps of the location of
$\triad$ in phase-space on approach to extinction, averaged for equal
time-to-extinction.  Far from extinction, there are small-amplitude
oscillations in densities, whenever stage two is synchronized (for
small $\alpha$, left), or none at all when stage two is a fixed point
(for large $\alpha$, right).  The amplitude of oscillations increases
shortly before extinction (black square), which occurs on the $RS$
line of the outer boundary, i.e.~when no more infected (I) sites
remain.

Notice that the amplitude of oscillations, early before extinction,
may be underrepresented in these plots. Different realizations of the
experiment, even if internally synchronized, are not in synchrony
\emph{with each other} too long before extinction. This is due to
slight differences in oscillation periods, which in turn produce phase
differences that are zero near extinction (because we make extinction
points coinciden when averaging), but are large and destroy
inter-sample coherence at early times.  However, consideration of
\Figs{fig:Fzsvstime1D.R} and \ref{fig:Fzsvstime2D.R}, where the
coherence $z$ is first measured for each sample and only later
averaged over samples, allows one to be certain that stage two has
nonzero coherence for small $\alpha$.
\subsection{Extinction as escape over a potential barrier}
\label{sec:extinction-as-escape}
The basic features of the extinction process, discussed in the
previous Section, are as follows. As \Figs{fig:Fzsvstime1D} and
\ref{fig:Fzsvstime2D} show, the persistence $F(t)$ stays at exactly
one during a ``resilience time'' $\tau_0$. There are no extinctions at
all during this period. Once the system has synchronized partially, on
stage two, a randomly occurring event triggers a sudden increase in
synchronization (the system enters stage three) and extinction happens
inevitably soon thereafter.
\begin{figure}[h!]
\centerline{
  \includegraphics[width=0.38\linewidth,angle=270]{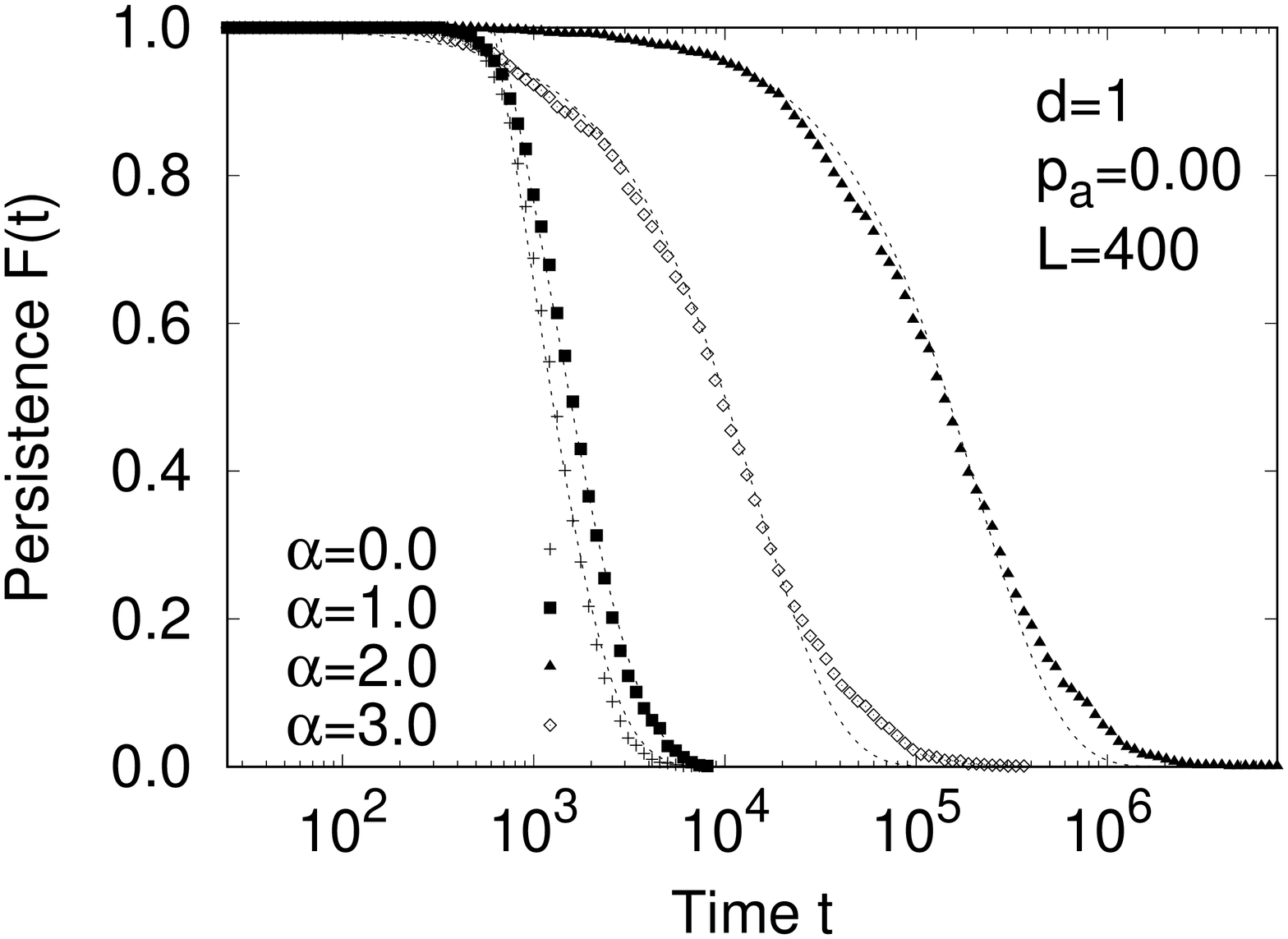}
  \includegraphics[width=0.38\linewidth,angle=270]{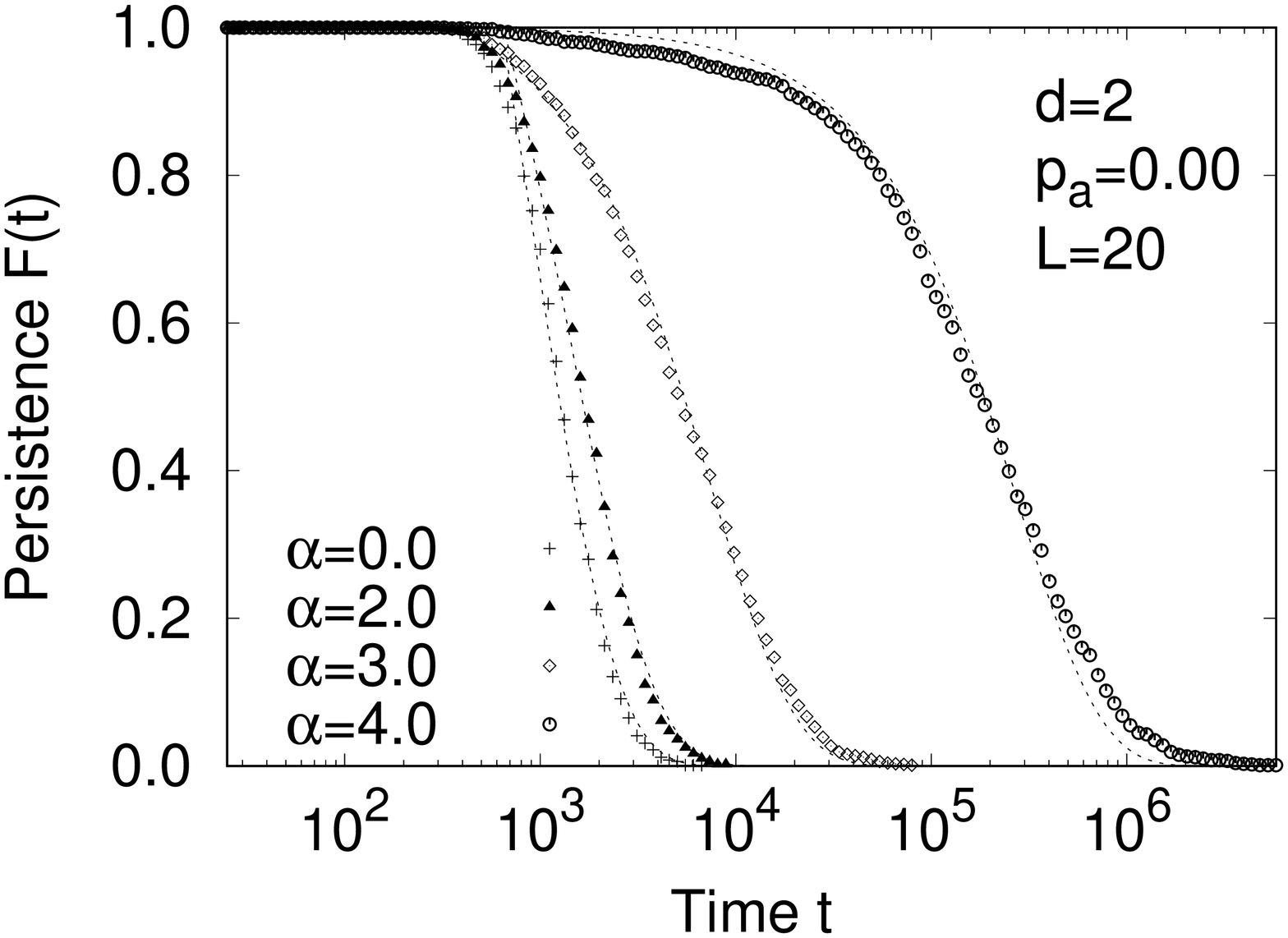}
}
\centerline{
  \includegraphics[width=0.38\linewidth,angle=270]{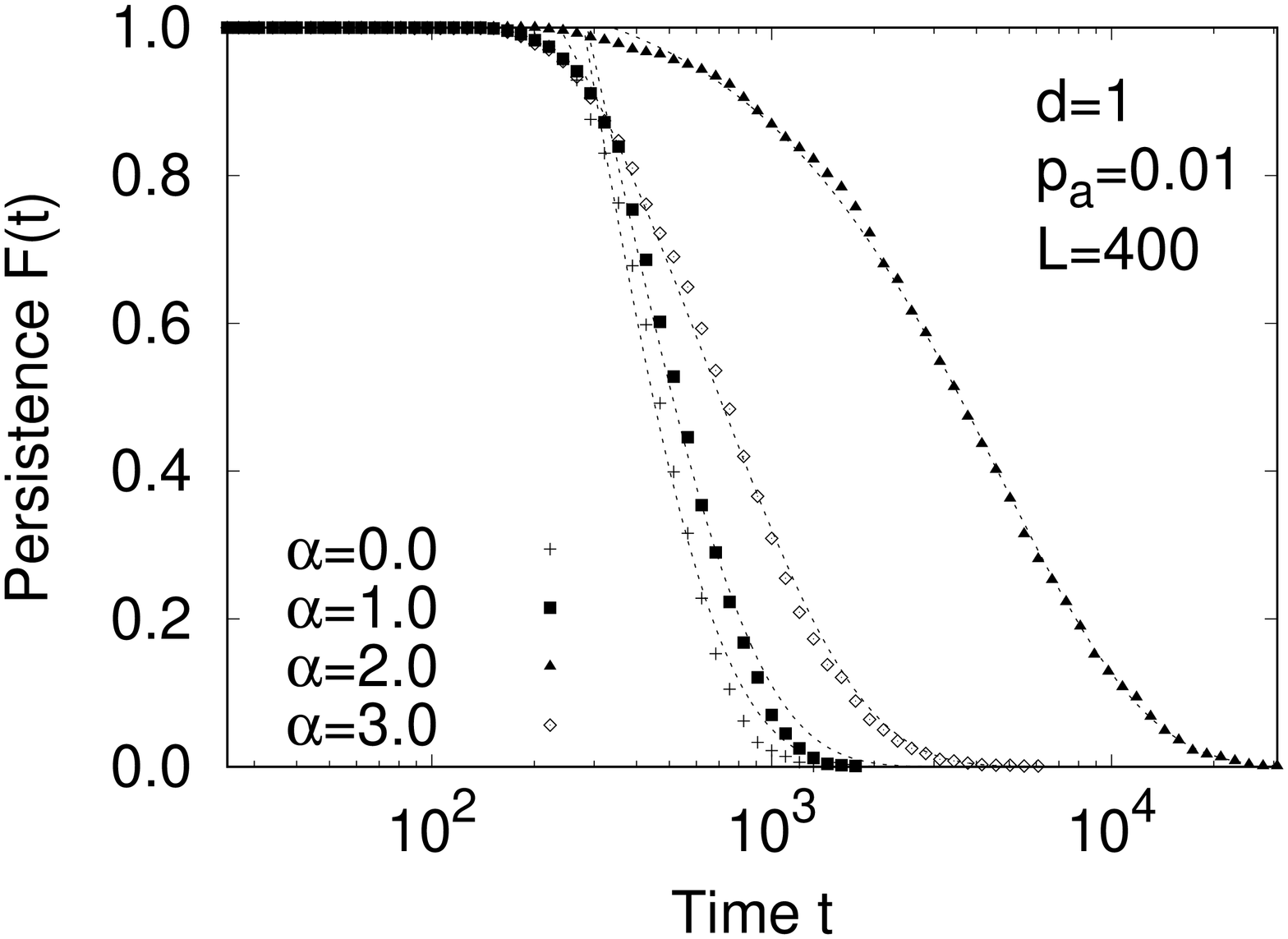}
  \includegraphics[width=0.38\linewidth,angle=270]{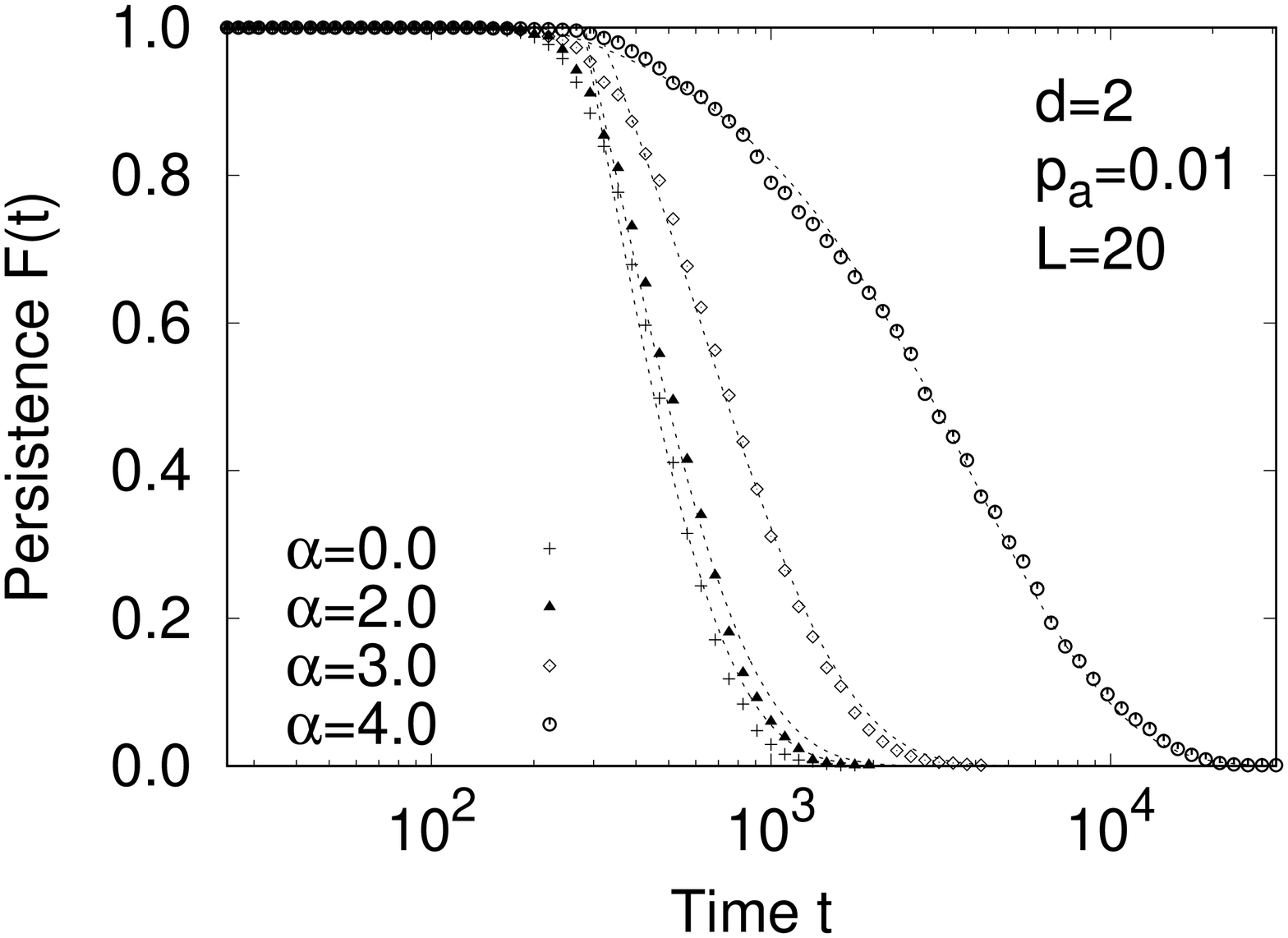}
}
\centerline{
  \includegraphics[width=0.38\linewidth,angle=270]{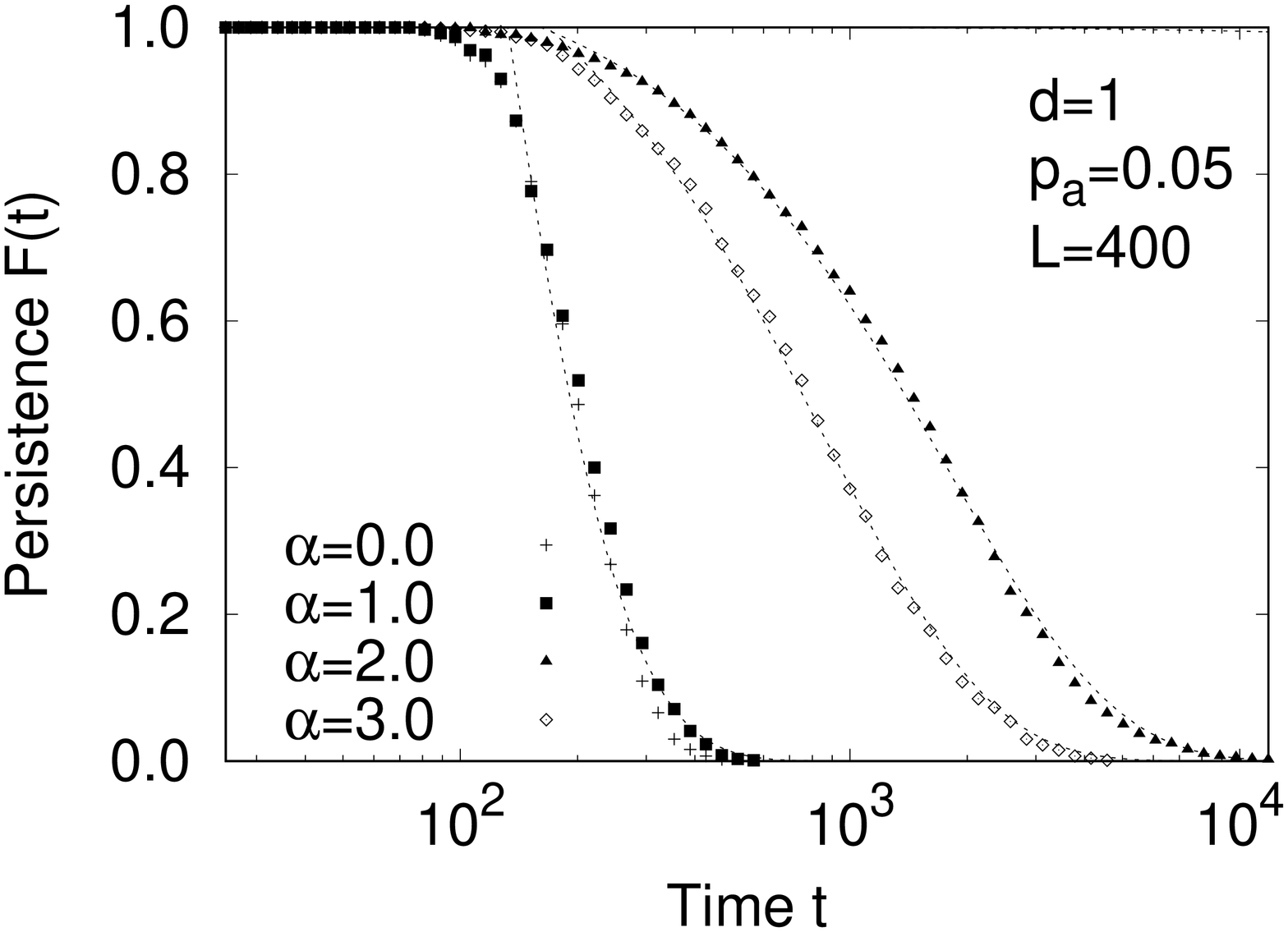}
  \includegraphics[width=0.38\linewidth,angle=270]{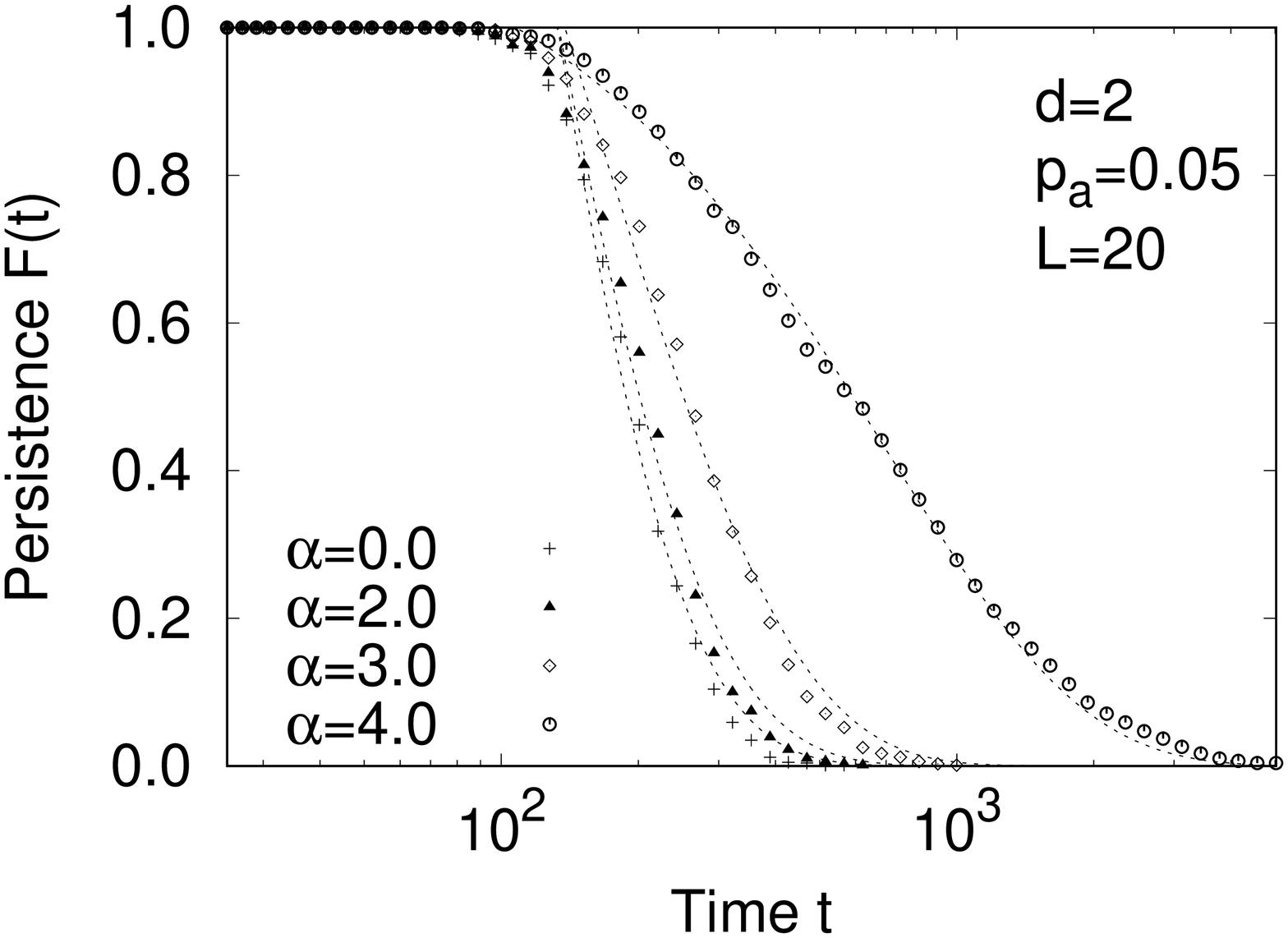}
}
\caption{Persistence $F(t)$ estimated as the fraction of samples still
  active at time $t$, for the SIRS model discussed in this work in one
  (left) and two (right) dimensions, for several values of the
  link-exponent $\alpha$ (symbol labels). In all cases, the number of
  sites in the system is $N=400$. After a ``resilient'' period
  $\tau_0$, during which no extinctions occur and $F(t)$ stays at one,
  $F(t)$ starts to decay and does so approximately exponentially in
  time.  Increasing the rate $p_a$ of link-annealing only shortens the
  involved timescales, without significantly modifying the above
  described dynamical behavior. Dashed lines are exponential fits
  using (\ref{eq:etimedist}), from which our estimates of $\tau_0$ and
  $\tau_2$ are obtained.}
\label{fig:persistence.vs.time}
\end{figure}

These observations can be rationalized by means of an approximate
treatment, as follows.  Consider the triad $\triad=(R,S,I)$ formed by
the fractions of Refractory, Susceptible, and Infected sites in the
system. Since $R+S+I=1$, this triad is constrained to a
plane. Non-negativity of fractions further constrains $\triad$ to the
interior of a triangle $\Omega$ within this plane (See
\Fig{phase.space.rev}).  A state of persistent asynchronous activity
(the \emph{endemic} state of an infection) is described, in this
representation, by a static point near the center of the available
domain.  A synchronized system, on the other hand, corresponds to a
point that evolves around a closed orbit. A trajectory that approaches
the external boundary $\partial \Omega$ of the available region
$\Omega$ takes the system close to dynamical extinction, as can be
seen by considering the time-evolution of a point on the RS line in
\Fig{phase.space.rev}). In the long run, all trajectories end up in
the absorbing state $S=1$.

\begin{figure}[h!]
\centerline{
  \includegraphics[width=0.70\linewidth,angle=270]{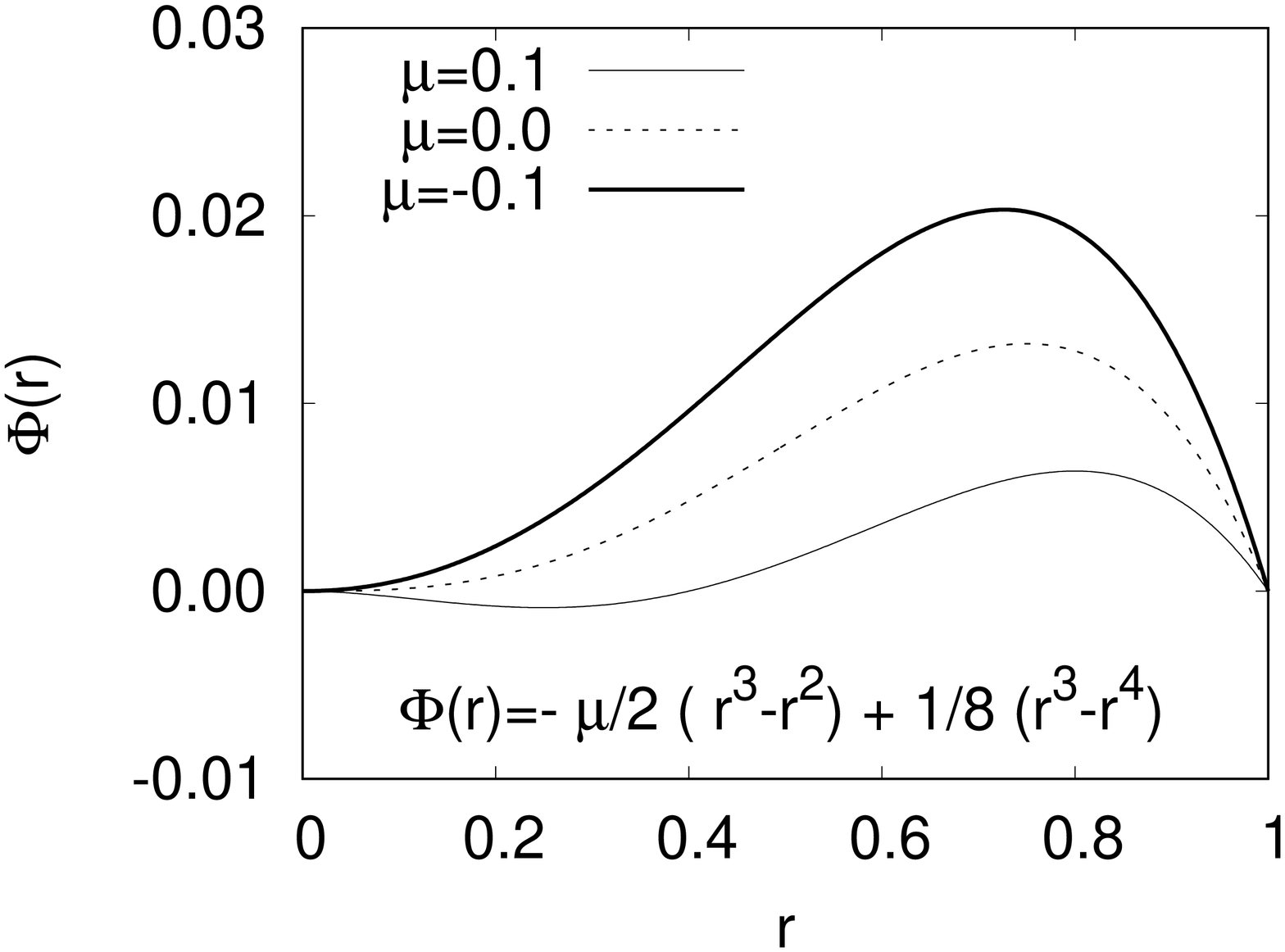}
}
\centerline{
  \includegraphics[width=0.45\linewidth]{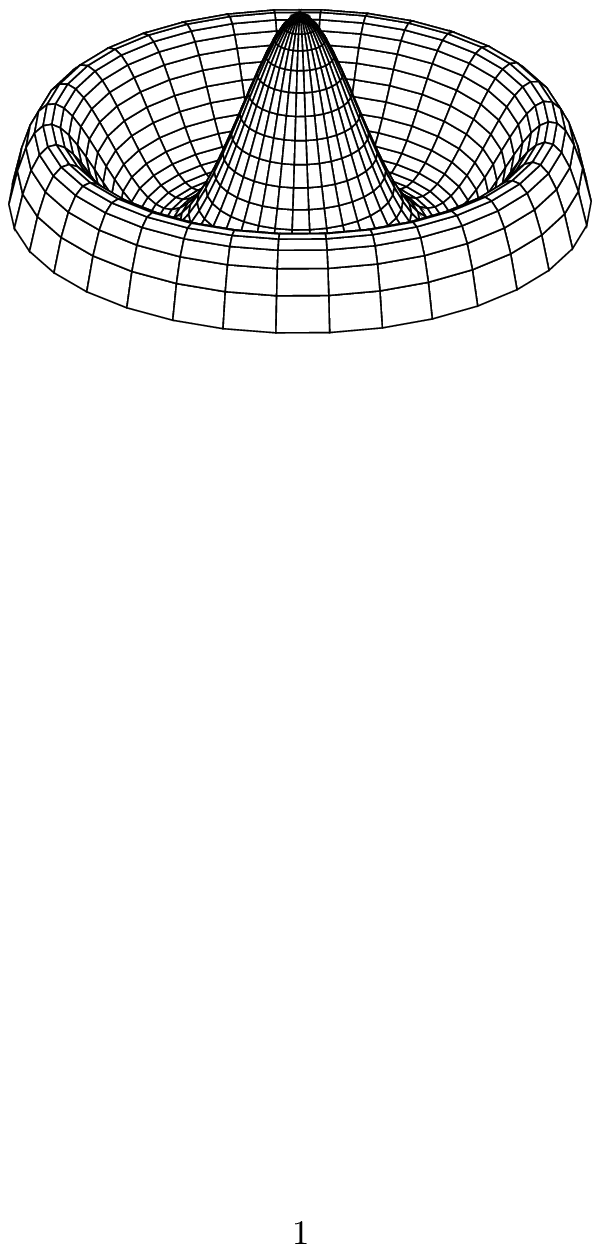}
  \includegraphics[width=0.45\linewidth]{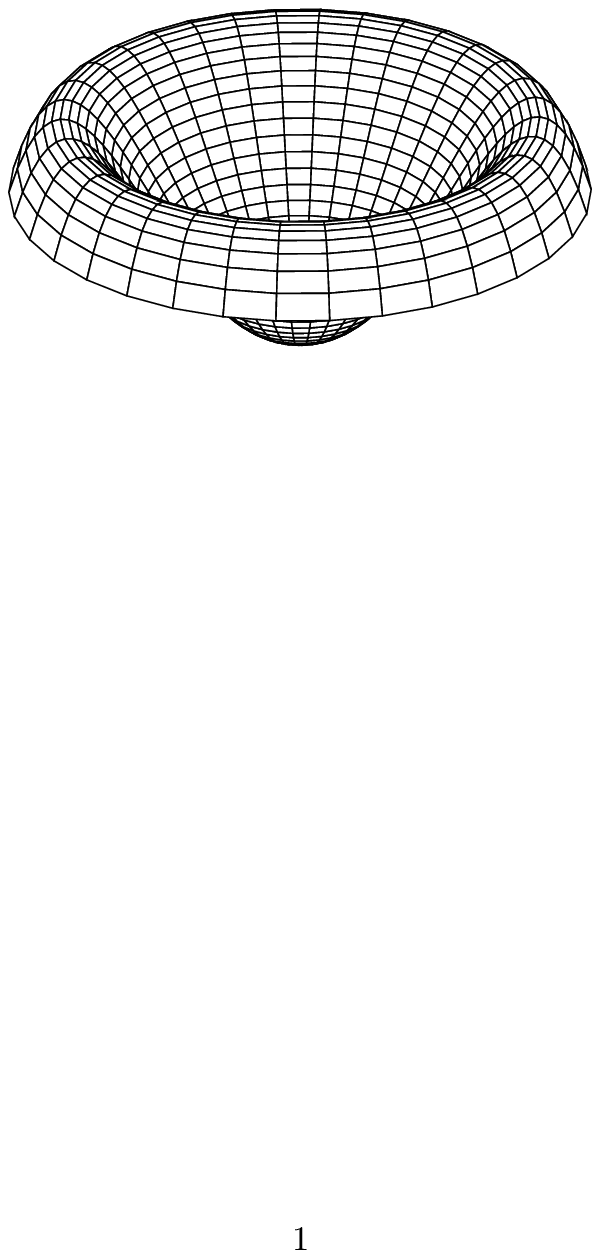}
}
\caption{Radial dependence (top) of the potential surface
  $\Phi(r,\theta)$ (bottom) associated with the observed dynamics,
  both for $\alpha<d$ (top, thin line, and bottom, left) in which case
  there is a stable periodic orbit, and for $\alpha>d$ (top, thick
  line, and bottom, right), when the endemic state at $r=0$ is
  stable. Notice that in both cases there is an unstable orbit (at the
  top of the potential barrier) engulfing the stable attractor, and
  beyond which the dynamics gets absorbed at the outer boundary. The
  case $\alpha=d$ or $\mu=0$ (top, dashed line) may correspond to a
  supercritical Hopf bifurcation, taking place inside an unstable
  orbit. }
\label{fig:mxhat}
\end{figure}

For the sake of simplicity, assume that periodic orbits, and the
external boundary $\partial \Omega$, are circular in shape, and that
the endemic equilibrium state, if there is any, is located at $r=0$,
which is also the random initial state. Next, and for the sake of
tractability, let us further assume that the \emph{noiseless} dynamics
of \triad can be written, in polar coordinates, as
\begin{eqnarray}
  \dot r &=& - \frac{\partial \Phi(r)}{\partial r} \nonumber  \\
  \dot \theta &=& \omega.
  \label{eq:2}
\end{eqnarray}
Actually, the last assumption, i.e.~that $\dot{\triad}$ is a
\emph{function of state}, is a rather strong one in this context. In
fact, for the SIRS model under consideration, the knowledge of
$\triad$ is clearly not enough to calculate $\dot{\triad}$, because
one also needs to know the time at which each site last became
infected. A rigorous treatment of the mean-field dynamics for large
systems~\cite{sirs-delay} requires the knowledge of the entire
distribution of times since last infection. The time since last
infection of course determines the present state of a node, but it
also provides more information. That extra information is needed in
order to calculate the noiseless dynamics. However, and for the sake
of mathematical tractability, we will relax that requirement, and
pretend that the sole knowledge of the density of nodes in each state,
i.e.~knowledge of \triad, suffices to calculate the dynamics, and that
the resulting equations are given by (\ref{eq:2}) for some potential
$\Phi(r)$.

Furthermore, on a frame that rotates with angular velocity $\omega$ we
would have
\begin{eqnarray}
  \dot r &=& - \frac{\partial \Phi(r)}{\partial r} \nonumber  \\
  \dot \theta &=& 0,
\end{eqnarray}
which is readily written as $\dot{\vec{\mathbf{x}}}= - \nabla
\Phi(\vec{\mathbf{x}})$ , where $\Phi$ is a rotationally invariant
potential. In the presence of noise, the dynamical equations in the
rotating frame read
\begin{equation}
\dot{\vec{x}}= - \nabla \Phi + \vec \xi,
\end{equation}
where $\vec \xi$ is a noise vector. Its angular component only
produces phase diffussion, while its radial component is responsible
for the eventual escape of the system from any existing stable
attractors of the noiseless dynamics.  For the purpose of representing
the dynamical phenomenology observed in our simulations, we will be
discussing the particular case of
\begin{eqnarray}
\Phi(r,\theta)= - {\mu}/{2} (r^3-r^2) + {1}/{8} (r^3-r^4),
\label{eq:3}
\end{eqnarray}
the properties of which are illustrated in \Fig{fig:mxhat}. For small
$\mu>0$, the endemic state at $r=0$ is unstable, and there is a stable
orbit located at $r^s(\mu)$, between $r=0$ and the outer unstable
orbit. This stable orbit attracts the noiseless dynamics. As $\mu \to
0^+$, the stable orbit radius $r^s(\mu)$ shrinks continuously to zero,
until this orbit coalesces with the unstable state at $r=0$ when
$\mu=0$. For small $\mu<0$, the endemic state at $r=0$ becomes a
stable attractor, while the external unstable orbit modifies its
radius only slightly. In the language of Dynamical
Systems~\cite{SNDAC94}, this transformation corresponds to a
\emph{Supercritical Hopf bifurcation}. In this particular case, one
occurring inside an unstable orbit. This simple potential reproduces
the observed dynamical stages if one further assumes that \hbox{$\mu
  \propto (d-\alpha)$}. Consequently, for $\alpha < d$ one has $\mu$
positive and therefore a stable periodic orbit surrounded by an
unstable one. For $\alpha > d$, on the other hand, $\mu$ is negative
and only the endemic state is stable, and it is still surrounded by an
unstable orbit. 

Let us now analyze the typical dynamical behavior of this system in
the presence of noise.  The radial coordinate $r$ is proportional to
the coherence and measures synchronization. The starting point is a
random mixture of $R,S$ and $I$, therefore non-synchronized and
located at $r=0$ within our simplified picture.  For $\alpha<d$, the
potential $\Phi$ has a local maximum at $r=0$ (thin line,
corresponding to $\mu= 0.1$ in \Fig{fig:mxhat} top, and bottom left
surface in that figure) . This state is unstable, so the system slides
downwards towards the stable orbit, in which constitutes stage one of
the dynamics. The period of variable duration spent at or near the
stable orbit is stage two. Eventually, a fluctuation pushes the system
over the external potential barrier, after which it enters stage
three. During stage three, synchronization increases rapidly until the
system gets absorbed at the external circular boundary.

When $\alpha>d$, the state $r=0$ is a minimum of $\Phi$ (thick line,
corresponding to $\mu= -0.1$ in \Fig{fig:mxhat} top, and bottom right
surface in that figure), and therefore locally stable. The system does
not synchronize initially, but fluctuates around the endemic state
($r=0$ in this case) until noise allows it to escape over the unstable
orbit, entering stage three to extinction. Significant synchronization
only occurs during stage three. Our system would thus undergo a
supercritical Hopf bifurcation~\cite{SNDAC94} at $\alpha=d$. For the
particular SIRS model studied here, this supercritical bifurcation is
located inside a passive unstable orbit, which gives rise to stage
three of the dynamics (rapid synchronization leading to extinction).

In addition to just providing a simple graphical illustration of the
origin of the observed three dynamical stages, this model makes very
specific scaling predictions that can be tested against numerical
results. In the presence of weak noise, and for $\alpha<d$, one
expects to see three distinct dynamical stages, when starting from a
state of random activity near the central peak of the potential: An
initial, downward moving stage of synchronization lasting for $\tau_1$
steps, followed by a synchronized stage of stochastic duration $t_2$,
and finally a third stage of increasing synchronization that leads to
extinction after $\tau_3$ steps. The durations $\tau_1$ and $\tau_3$
of stages one and three are expected to be well defined (not
stochastic) if the amount of noise is small, because during these
stages the dominant forces are the deterministic ones. But the
permanence time $t_2$ of the system stage two, being related to a
process of noise-induced escape against deterministic forces, should
be strongly stochastic.

The distribution of $t_2$ and the scaling with $N$ of the involved
timescales is discussed in the following Sections, and compared with
predictions stemming from the noise-activated-escape picture discussed
above.
\subsection{Exponential distribution of permanence times in stage two}
\label{exp-dist-t2}
Using concepts from the theory of First-Passage
processes~\cite{RGFPP01}, one can conclude that, due to the finite
character of the segment in which the radial variable executes its
escape process, the distribution of escape times from the minimum of
the potential must be asymptotically exponential, with a dominant
timescale that we denote by $\tau_2$.

Before an extinction can take place, the system must at least traverse
stages one and three, which have a combined duration, of
$\tau_0=\tau_1+\tau_3$ timesteps.  Therefore, the persistence $F(t)$
has to be rigorously equal to one during a ``resilience period'' of
$\tau_0$ timesteps, and it has to decay exponentially for long
times. To first approximation, one can expect a functional form like
\begin{equation}
  \label{eq:etimedist}
  F(t) = \theta(\tau_0-t) + \theta(t-\tau_0) e^{(t-\tau_0)/\tau_2},
\end{equation}
where $\theta$ is the Heavyside step function, and
$\tau_0 = \tau_1+\tau_3$ is the added length of stages one and three. The
permanence time $t_2$ of the system in stage two is a stochastic
variable with an exponential distribution characterized by a unique
parameter $\tau_2$.

We estimate $\tau_0$ and $\tau_2$ by fitting (\ref{eq:etimedist}) to
our data for the persistence $F(t)$, in 1d and 2d, for
\hbox{$p_{a}=0.00, 0.01, 0.05$} and for several system sizes and
values of $\alpha$. A subset of these data are shown in
\Fig{fig:persistence.vs.time} toghether with their corresponging
fits. The resulting estimated values of $\tau_0$ and $\tau_2$ are
presented in \Figs{fig:tau0vsl} and \ref{fig:tau2vsl}.
\subsection{Scaling of resilience time $\tau_0$ with system size $N$}
\label{sec:scal-refr-time}
\begin{figure}[h!]
\centerline{
\includegraphics[width=0.50\linewidth]{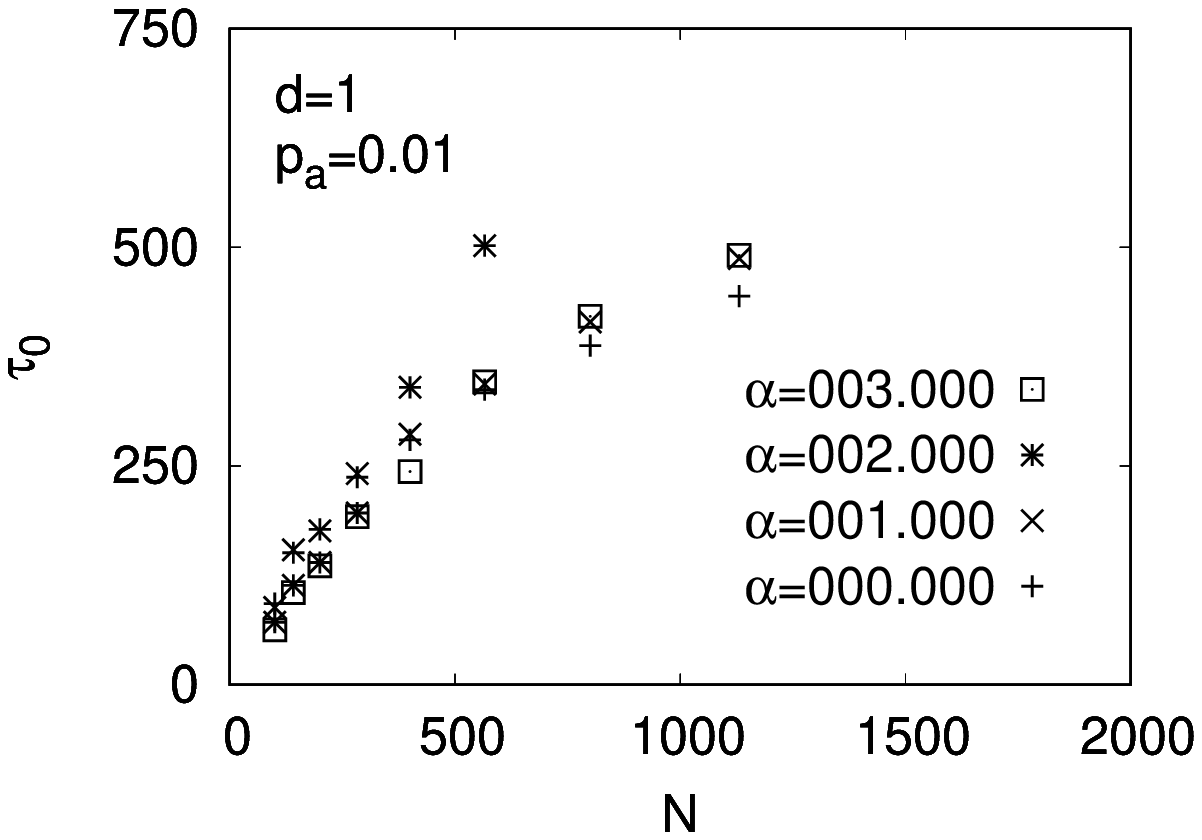}
\includegraphics[width=0.50\linewidth]{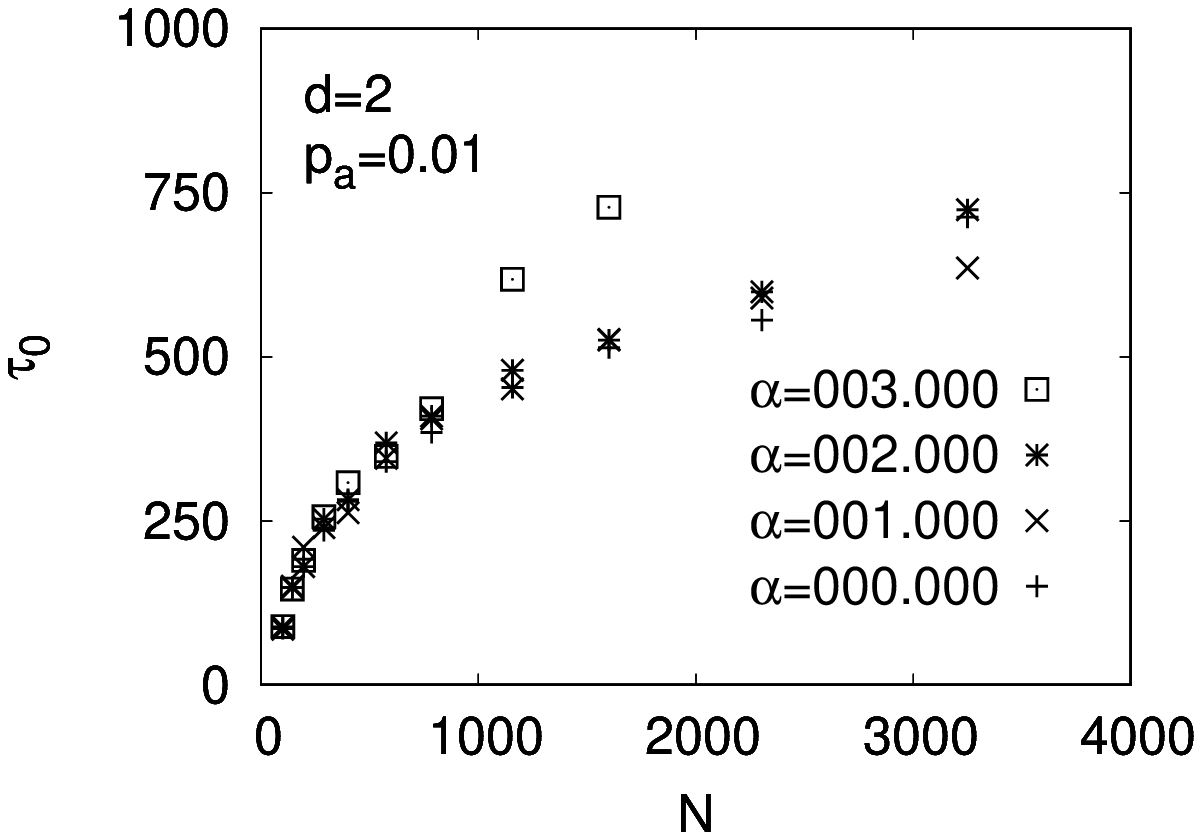}
}
\caption{Scaling with system size $N$, of the ``resilience time''
  $\tau_0$ during which the persistence $F(t)$ stays at one, in one
  (left) and two (right) dimensions, for some values of the
  link-length exponent $\alpha$. Theoretical considerations discussed
  in the text make one expect a logarithmic scaling of the form
  $\tau_0 \sim \log(N)$, which is consistent with the behavior seen in
  these plots.}
\label{fig:tau0vsl}
\end{figure}
Results for the estimated values of the resilience time $\tau_0$,
obtained from fits of (\ref{eq:etimedist}) to our data, are presented
in \Fig{fig:tau0vsl}.  As seen in these figures, our results are
consistent with a logarithmic growth of the type \hbox{$\tau_0 \sim
  \log{N}$}.

According to the mental model depicted in \Fig{fig:mxhat}, the
resilience time $\tau_0$ is the time it takes for the system to slide
downwards from the starting state onto the stable orbit (in case one
exists), plus the time from the ridge to the outer boundary, where
extinction takes place. Both are diffusive motions along a flow, and
therefore a logarithmic scaling with $N$ is
expected~\cite{exit-problem-grasman99}. This expectation is well
verified by the data displayed in \Fig{fig:tau0vsl}.
\subsection{Scaling of escape timescale $\tau_2$ with system size $N$}
\label{tau2scaling}
\begin{figure}[h!]
\centerline{
\includegraphics[width=0.50\linewidth]{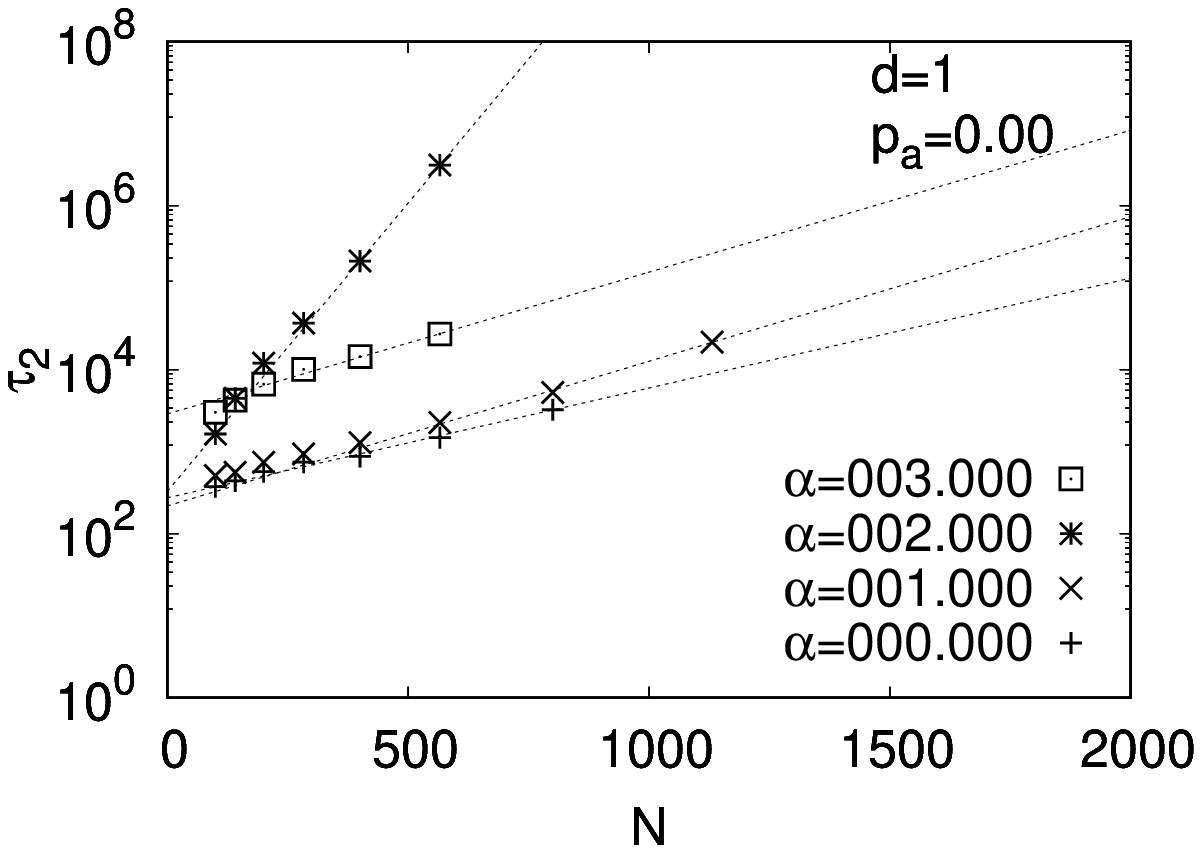}
\includegraphics[width=0.50\linewidth]{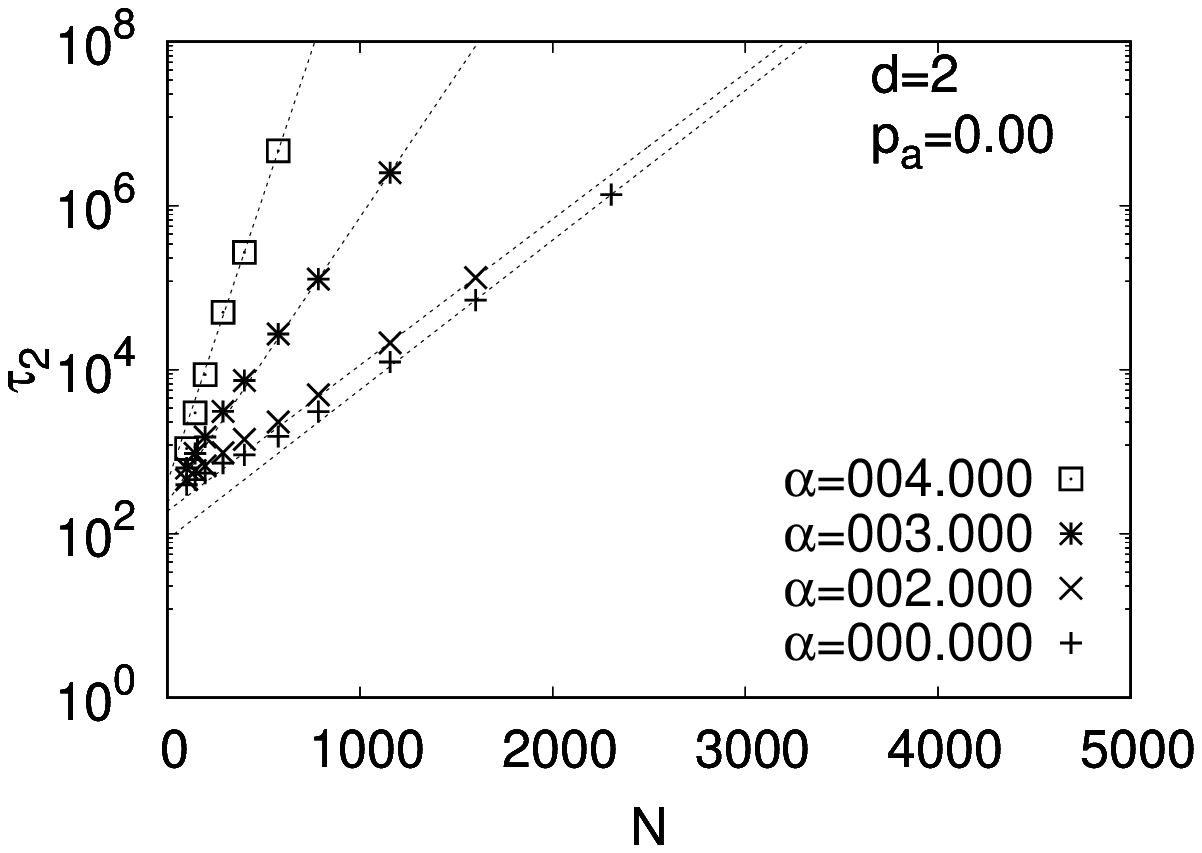}
}
\centerline{
\includegraphics[width=0.50\linewidth]{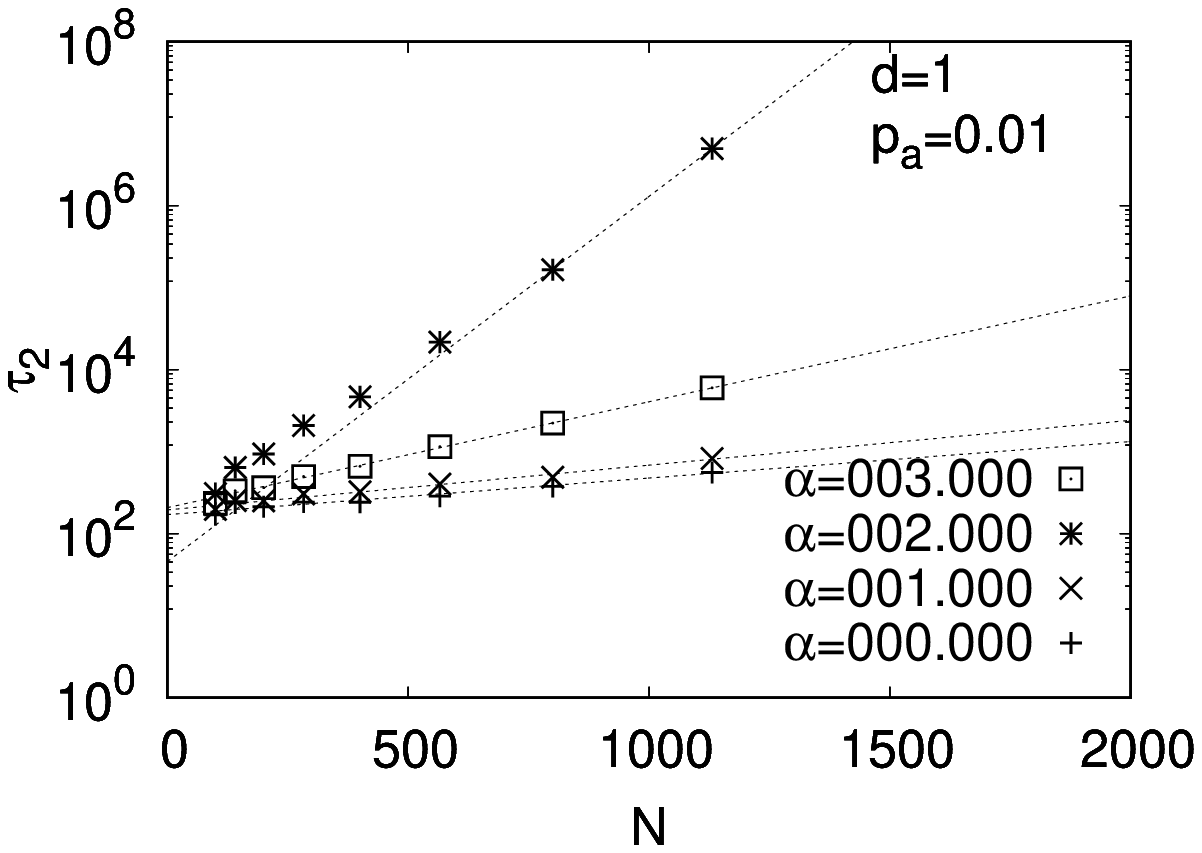}
\includegraphics[width=0.50\linewidth]{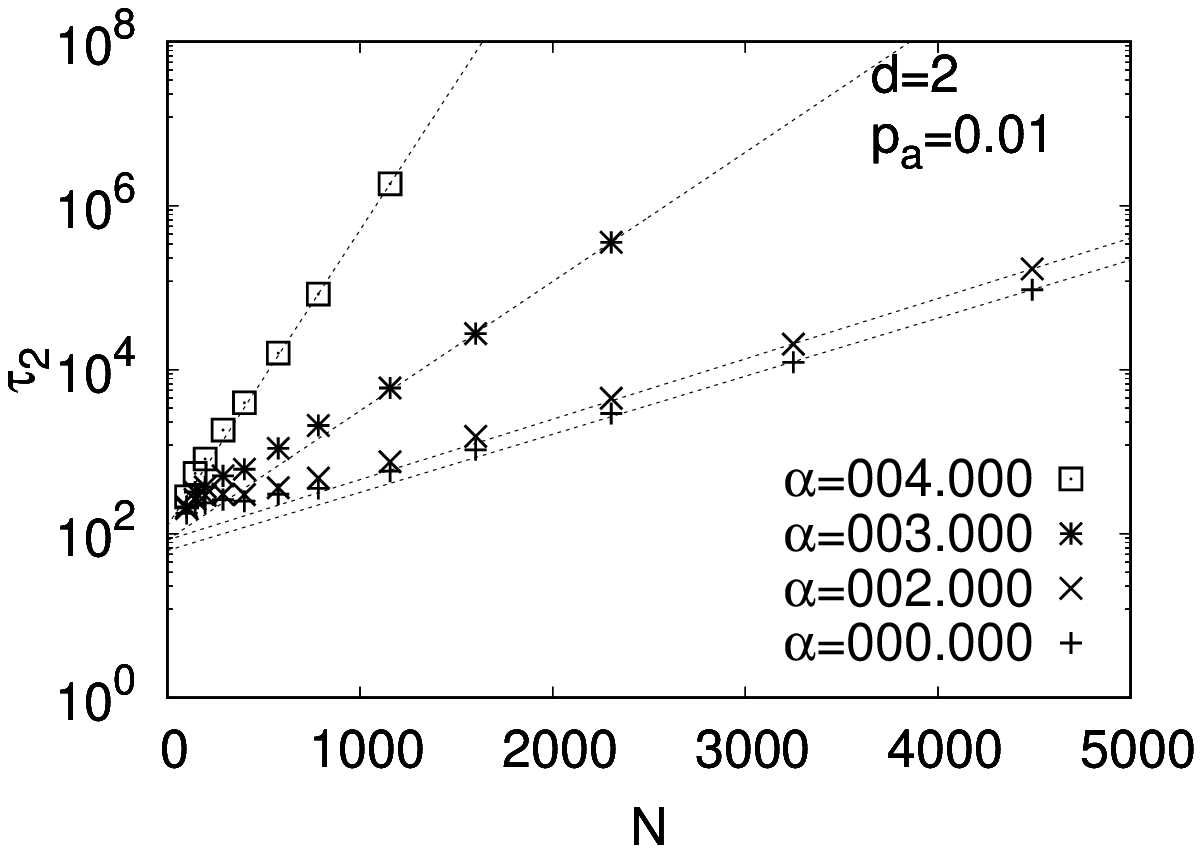}
}
\centerline{
\includegraphics[width=0.50\linewidth]{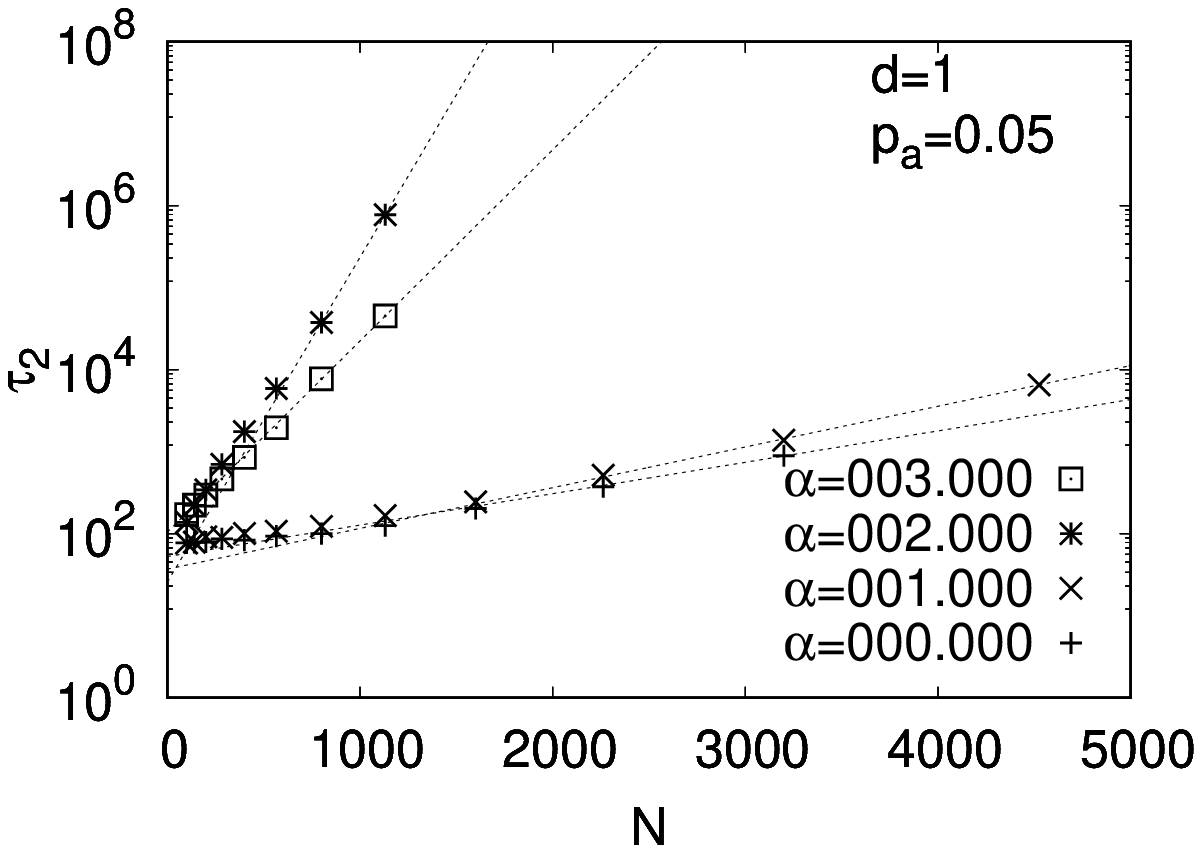}
\includegraphics[width=0.50\linewidth]{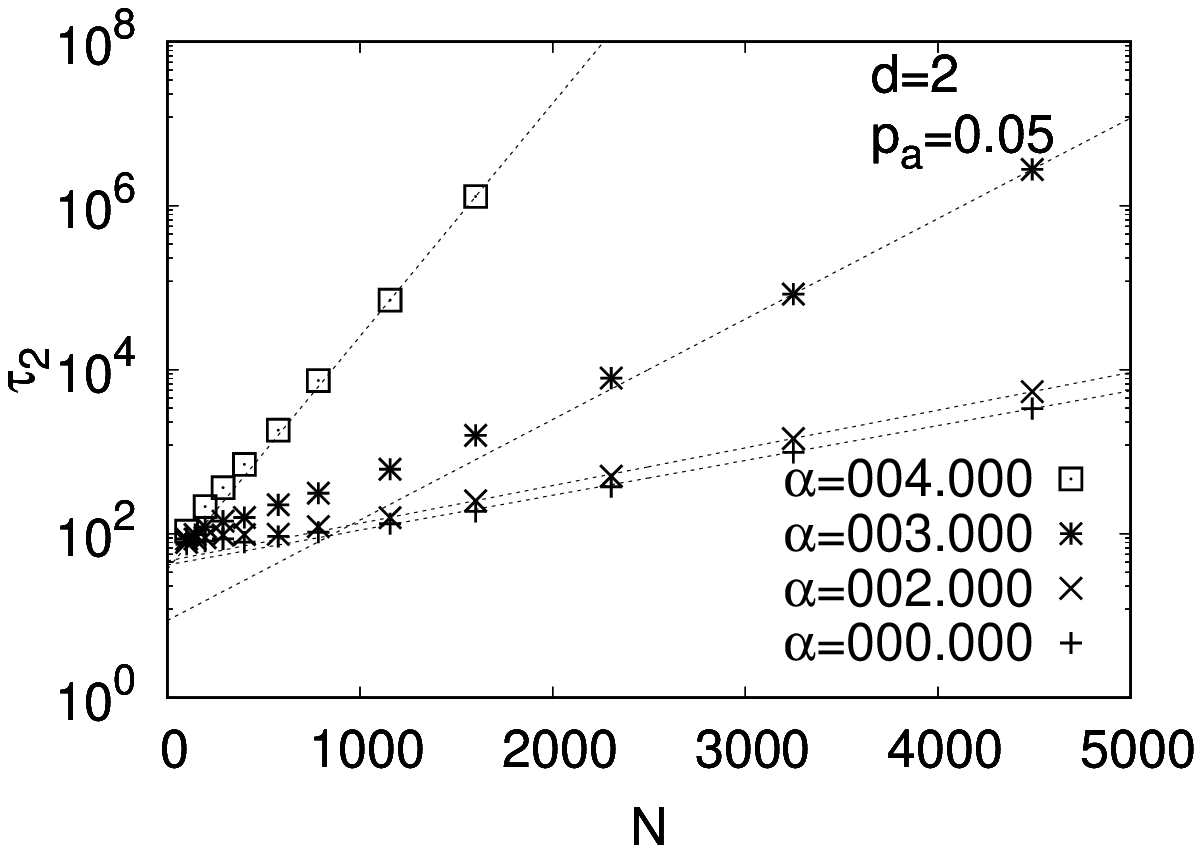}
}
\caption{Timescale $\tau_2$ for permanence in stage two (See
  (\ref{eq:etimedist})) versus system size $N$ in one (left) and two
  (right) dimensions, for three values of the network-annealing
  parameter $p_a$, and for several values of the link-length parameter
  $\alpha$.  An asymptotically exponential growth of $\tau_2$ with $N$
  is seen in these plots. }
\label{fig:tau2vsl}
\end{figure}

The escape process from the stable orbit to the ridge is, in the
scenario of \Sec{sec:extinction-as-escape} and \Fig{fig:mxhat}, a
noise-induced escape against a flow, that is, escape over a potential
barrier. In this case, $\log{(\tau_2 N^{1/2})}$ is expected to scale
linearly with system size
$N$~\cite{extinction-sir-removal-renewal}. Asymptotically, one would
therefore expect $\tau_2 \sim e^{\kappa_2 N}$ to hold.

Our results for the estimated values of $\tau_2$, presented in
\Fig{fig:tau2vsl}, are in fact consistent with the expectation that,
for large $N$, the logarithm of the dominant timescale $\tau_2$ for
permanence in stage two increases linearly with $N$, thus confirming
the abovementioned expectations.

\subsection{Speed of growth $\kappa_2$ of  $\tau_2$ with size $N$}
\label{sec:speed-growth}
\begin{figure}[h!]
  \centerline{
    \includegraphics[width=0.38\linewidth,angle=270]{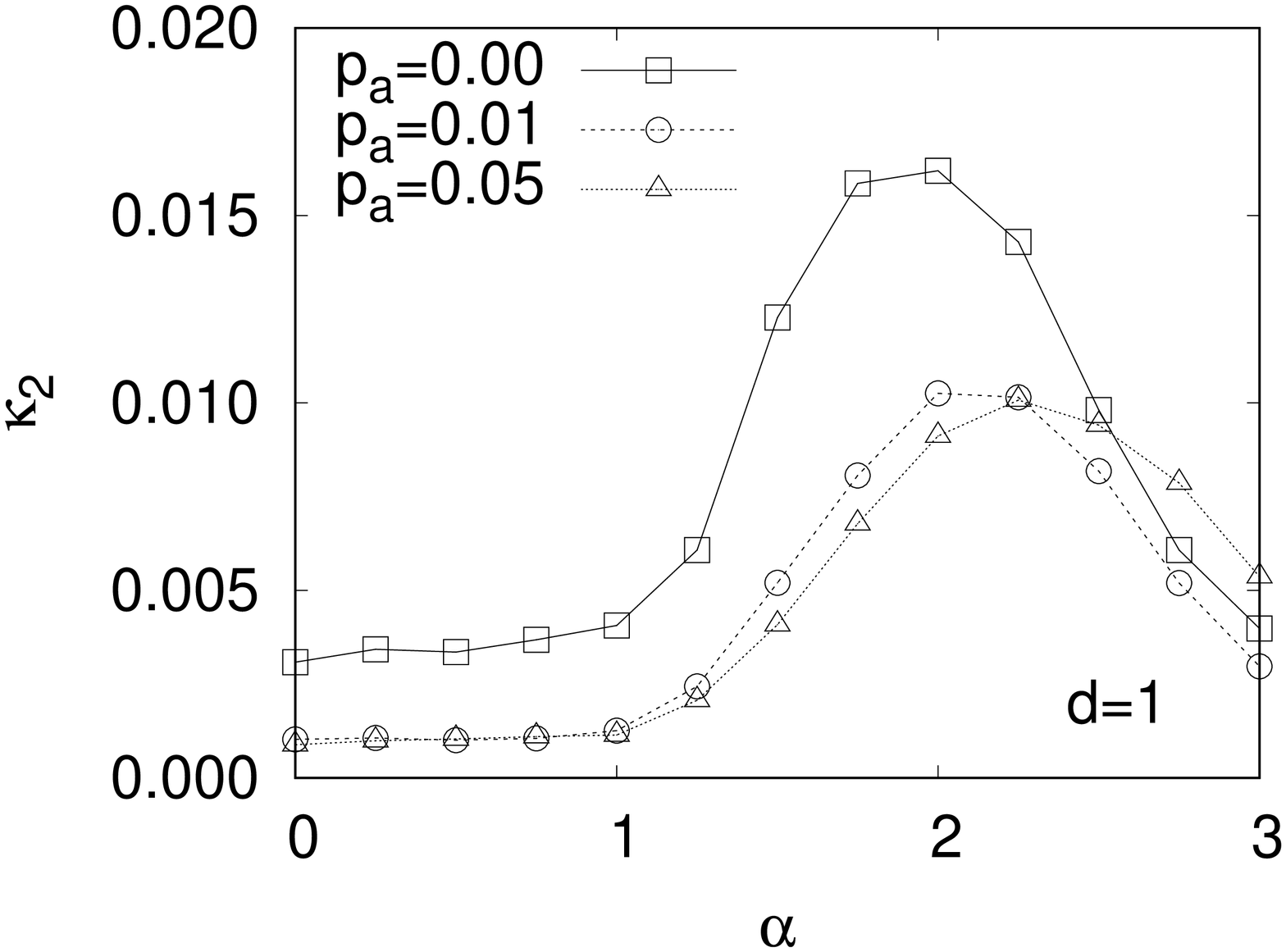}
    \includegraphics[width=0.38\linewidth,angle=270]{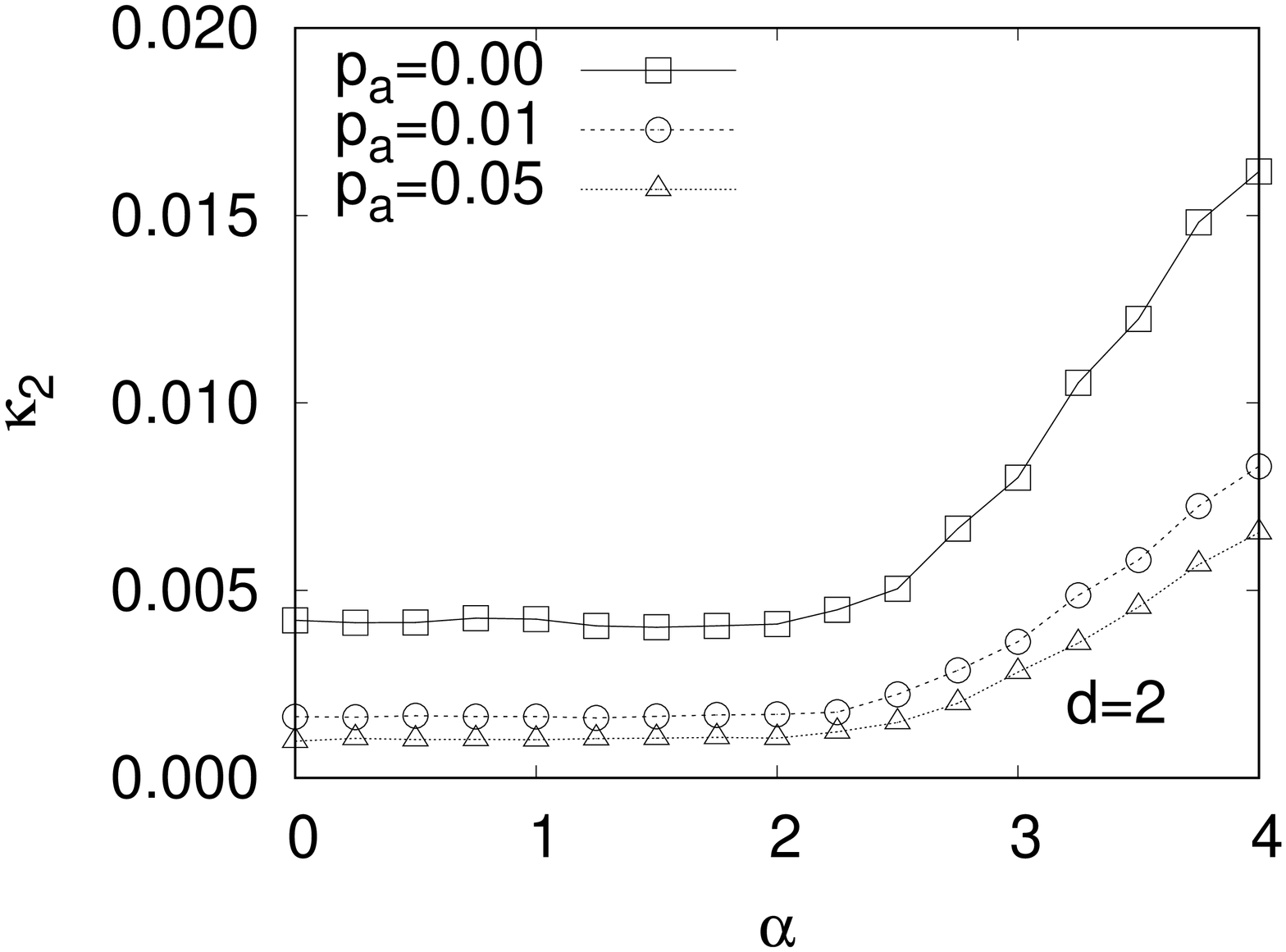}
  }
  \caption{Logarithmic speed of growth $\kappa_2$ vs size $N$ of the
    timescale $\tau_2$ for escape from stage two of the dynamics. In
    one dimension (left) there is a maximum in $\kappa_2$ near
    $\alpha=2d=2$, which is not seen in two-dimensions (right). }
  \label{fig:kappa2.vs.alpha}
\end{figure}

As discussed in previous sections, our results show that $\tau_2 \sim
e^{\kappa_2 N}$ asymptotically. We determine $\kappa_2(d,\alpha)$ as
the slope of the asymptotic linear fits shown in
\Fig{fig:tau2vsl}. Our results are presented in
\Fig{fig:kappa2.vs.alpha} in one and two dimensions, for three values
of the annealing probability $p_a=0,0.01$ and $0.05$ used in this
work. In one dimension, $\kappa_2$ is found to have a maximum, for any
amount of annealing, near $\alpha \approx 2$. This hints at some kind
of dynamical phase transition, since escape times appear to diverge
much faster with system size in the neighborhood of $\alpha=2$. This
maximum is apparently not observed in two dimensions, where we have
explored a few instances with large values of $\alpha$ (up to
$\alpha=8$), without seeing a decrease in escape times as the one
observed in one dimension. We cannot discard, however, that a similar
nonmonotonic behavior of $\kappa_2(\alpha)$ occurs in two dimensions
as well, for larger systems and/or values of $\alpha$, although this
does not seem probable.  Furthermore, more work would be needed in
order to clarify the precise meaning of the maximum in $\kappa_2$ that
is observed in one dimension near $\alpha=2$.
\section{Conclusions and Discussion}
\label{sec:concl-disc}
We analyzed the process leading to dynamical extinction in a
high-infectivity SIRS model with fixed infectious ($\tau_{i} = 4$) and
refractory ($\tau_{r} = 8$) periods, on long-range connected networks
of $N$ sites in one and two dimensions. The direct-link connectivity
probability $p_{ij}$ for an arbitrary pair of sites $i$ and $j$ at a
distance $r_{ij}$ from each other, decays as
\hbox{$1/r_{ij}^{\alpha}$}. There are two links per site, which
ensures that most sites belong to the largest connected component.

We focused on the synchronization and extinction properties of these
excitable systems, and their dependence on $N$ and $\alpha$. Starting
from a random mixture of Refractory, Infected, and Susceptible sites,
when $\alpha<d$ the system first undergoes partial synchronization and
then, after a varying amount $t_2$ of time, the coherence increases
even further and the dynamics becomes extinct (the number of infected
sites becomes zero). The link infectivity is $p_0=0.75$, large enough
to ensure that spontaneous extinction of the dynamics, not mediated by
synchronization, is not relevant~\cite{sirs-sw-ka}.

In addition to static networks with a fixed set of links, we also
analyzed annealed ones. For these, at each timestep, each link is
replaced by an equally distributed one with probability $p_{a}$. In
this work we used three values of annealing: \hbox{$p_a=0.00, 0.01,
  0.05$}. We find that annealing only shortens the dynamical
timescales involved in the synchronization and extinction processes,
while keeping all static results quantitatively unmodified.

In \Sec{section:simulation-results} we argue that for $\alpha<d$, the
dynamical extinction process can be decomposed into three dynamical
stages that are described as follows: First a short period of rapid
synchronization starting from a random state. This first stage lasts
for approximately $10^3$ timesteps for the sizes studied here (See
\Figs{fig:Fzsvstime1D} and \ref{fig:Fzsvstime2D} ). Once the system is
dynamically organized and displays sustained oscillations, it stays in
this ``partially synchronized'' second stage for a (large) random
amount of time. At a random time while in stage two, a sudden increase
in synchronization happens, spanning approximately $10^3$ steps (See
\Figs{fig:Fzsvstime1D.R} and \ref{fig:Fzsvstime2D.R}), which we call
Stage three, and which inevitably leads to the extinction of the
dynamics. When $\alpha >2$, on the other hand, stage one does not
existe, stage two is a non-synchronized endemic state lasting for a
random amount of time, after which stage three happens as described
above.

In \Sec{sec:extinction-as-escape} we showed that all three stages of
the dynamics have an interpretation in terms of the potential depicted
in \Fig{fig:mxhat}, in the presence of noise. This ``potential'' is
assumed to depend on two space coordinates of the plane in which the
triad $\triad=(R,S,I)$, describing the state of the system, evolves in
time. In a simplified picture that approximates periodic orbits as
being circular, one can consider polar coordinates in this plane, with
their origin in the center of the periodic orbit.  The angular
coordinate $\theta$ describes the periodic motion of the synchronized
system as it cycles aroung its orbit. The radial coordinate $r$ is a
measure of the intensity of synchronization, and can be taken to be
proportional to the coherence $z$ (See \Eqn{eq:1}). In
\Sec{sec:extinction-as-escape}, the noiseless dynamics for $\triad$ is
argued to be approximately determined by the gradient of a potential,
while noise adds a small amount of diffusive behavior on top of
that. In the absence of noise, and when starting from a mixed state
(the central peak), if $\alpha<d$ the system would be attracted by,
and execute a periodic motion on, the stable orbit, which is a local
minimum of the potential. Because of noise, however, $\triad$ makes
random excursions away from its stable orbit, until it happens to jump
over the barrier and falls outwards, where synchronization attains
extreme values and the dynamics becomes extinct at the outer
boundary. The extinction process is, in this picture, described as
noise-activated escape over a potential barrier. When $\alpha>d$ the
stable orbit shrinks to zero size, rendering the endemic state $r=0$
stable. In that case, escape occurs over the potential barrier, from a
fixed point of the dynamics located at $r=0$.

This picture, which borrows from usual approaches in the stydy of
dynamical systems~\cite{SNDAC94}, provides a simple explanation for
the existence of three dynamical stages, but also imposes very
specific constraints on the scaling of permanence times in the
different stages with system size $N$.  Previous work regarding
diffusive motion in the presence of a deterministic flow predicts that
the combined time $\tau_0$ spent in stages one and three, being these
cases of motion along a flow, is only weakly affected by noise and
should scale with system size also weakly, i.e~as \hbox{$\log N$}. The
permanence time in stage two, on the other hand, depends on a strongly
stochastic diffusive process against a flow, that is, a noise
activated escape over a potential barrier. The timescale associated
with this escape process should then scale as $e^{\kappa_2
  N}$. Additionally, the distribution of permanence times in stage
two, according to very general arguments applying to first-passage
processes, is expected to be asymptotically exponential for long
times. All these predictions for the involved timescales are very well
verified using the results of our numerical experiments, as discussed
in \Secs{exp-dist-t2}, \ref{sec:scal-refr-time}, and
\ref{tau2scaling}.

Our measurements of $z_{1/2}$, the coherence at the half-time for
extinction, which are shown in
\Fig{fig:coherence.at.halftime.vs.alpha}, strongly suggest that stage
two has nonzero coherence $z$ in the thermodynamic limit, only if
$\alpha < d$. For larger values of $\alpha$, there is some small
amount of synchronization during stage two for finite systems, but it
becomes smaller as $N$ grows.

Although more work is needed to confirm the precise critical value of
$\alpha$ in the light of previous
work~\cite{entrainment-lower-critical-dimension,sakaguchi-kuramoto-model,synch-in-spatial-extended-oscillators,long-range-km-spin-wave},
it is clear that, for large $\alpha$, no state of sustained
synchronization is expected to exist on large systems. Stage one is
therefore not relevant in this case, and stage two becomes a state of
non-synchronous persistent activity in which the system lingers for a
long time. Then the system enters Stage three, in which explosive
synchronization, followed by extinction, occurs.

If the proposed picture of a supercritical Hopf bifurcation at
$\alpha=d$ is correct, then the radius of stable orbit below
$\alpha=d$, shrinks to zero continuously at $\alpha=d^-$.  In this
case, the $\alpha$-driven transition in $z_{1/2}$ that is seen in
\Fig{fig:coherence.at.halftime.vs.alpha} is a continuous
transition. The data in \Fig{fig:coherence.at.halftime.vs.alpha},
however, could as well be compatible with a discontinuity in $z_{1/2}$
at $\alpha=d$.  If this were the case, then the bifurcation at
$\alpha=d$ would be a \emph{saddle-node or fold} bifurcation (Sec. 8.4
in~\cite{SNDAC94}) instead of a supercritical Hopf bifurcation, as
assumed here. These two possibilities cannot be clearly distinguished
on the basis of our present data. Clearly, much more work would be
needed to establish the character (continuous or discontinuous) of
this $\alpha$-driven transition precisely.

The fact that there is no synchronized stage two for large $\alpha$,
poses the question of how this modifies the picture discussed above
and in \Sec{sec:extinction-as-escape}, that describes extinction as
escape over a barrier. The answer is that this picture is essentially
the same in the case of large $\alpha$, the only difference being that
the escape process occurs now starting from a single local minimum at
$z=0$, instead as from a stable orbit with finite $z$. The
distribution of escape times and their scaling with size, however, are
expected to be the same for escape from a local minimum at large
$\alpha$ as for escape from a stable orbit for small $\alpha \leq d$.

One of the few differences we find between one and two-dimensional
systems concerns the $\alpha$-dependence of $\kappa_2(\alpha,d)$ (See
\Fig{fig:kappa2.vs.alpha}), which measures the speed of growth of
$\log \tau_2$ with system size $N$.  As seen in
\Fig{fig:kappa2.vs.alpha}, $\kappa_2$ has a maximum near $\alpha =2$
for one dimensional systems. This appears to be an indication of a
dynamical phase transition, the origin of which we do not discuss in
this work. In two dimensions, on the other hand, we find that
$\kappa_2$ grows with increasing $\alpha$ when $\alpha > 2$ until at
least $\alpha=4$. Large values of $\alpha$ are not easy to simulate
because of long extinction times. However, we did some simulations
with $p_a=0.05$ (annealing shortens timescales) for $\alpha$ up to 8,
without finding any sign of a nonmonotonic behavior of
$\kappa_2(\alpha)$.

\section*{Acknowledgments}
The authors acknowledge the use of computational resources in clusters
Xiuhcoatl(CGSTIC) and Kukulkan (Dept. of Applied Physics) of
CINVESTAV. EAM thanks CONACYT for support through a PhD fellowship.
\bibliography{refs.bib} \bibliographystyle{unsrt}
\end{document}